\documentclass[symmetry,article,accept,pdLaftex,moreauthors]{Definitions/mdpi} 

\firstpage{1} 
\pubvolume{1}
\issuenum{1}
\articlenumber{0}
\pubyear{2024}
\copyrightyear{2024}
\externaleditor{\textls[-25]{Academic Editor: Firstname Lastname}}
\datereceived{7 September 2024 } 
\daterevised{9 October 2024 } % Comment out if no revised date
\dateaccepted{14 October 2024 } 
\datepublished{ } 
\hreflink{https://doi.org/} % If needed use \linebreak
\Title{A Survey of Dynamical and Gravitational Lensing Tests in Scale Invariance: The Fall of Dark Matter? %MDPI: Title is different from the one submitted in our system. Please check and confirm the correct one.
}

\TitleCitation{A Survey of Dynamical and Gravitational Lensing Tests in Scale Invariance: The Fall of Dark Matter?}

\Author{Andr\'e Maeder %MDPI: Please carefully check the accuracy of names and affiliations.
 $^{1,}$* %MDPI: Corresponding symbol was added according to the information submitted in our system. Please check and confirm.
\orcidA{}, Fr\'ed\'eric Courbin $^{2,3}$\orcidB{}}

\AuthorNames{Andr\'e Maeder, Fr\'ed\'eric Courbin}

\AuthorCitation{Maeder, %MDPI: Please carefully check the accuracy of names and affiliations.
 A.; Courbin, F.}
\address{%
$^{1}$ \quad Department of Astronomy, University of Geneva, Chemin des Maillettes 51, CH-1290 Sauverny, Switzerland\\
$^{2}$ \quad \textls[-15]{ICC-UB Institut de %MDPI: Please use English for institution types if possible. Same for university below.
 Ci\`encies del Cosmos, Universitat de Barcelona, Mart\'i Franqu\`es, 1, E-08028 Barcelona, Spain; frederic.courbin@icc.ub.edu} \\
$^{3}$ \quad ICREA, %MDPI: Please provide full name of istitution, not abbreviation if possible.
 Pg. Llu\'is Companys 23, E-08010 Barcelona,  Spain
}

 \corres{Correspondence: andre.maeder@unige.ch  }

\abstract{We first briefly review the adventure of scale invariance  in physics, from Galileo Galilei, Weyl, Einstein, and Feynman to the revival by Dirac (1973) and Canuto et al. (1977). In the way that the geometry of space--time can be described by  the coefficients $g_{\mu \nu}$, a gauging condition given by a scale factor $\lambda(x^{\mu})$ is needed to express the  scaling. In general relativity (GR), $\lambda=1$. The ``Large Number Hypothesis'' was taken  by Dirac and by  Canuto et al. to fix $\lambda$. The condition that {\emph{the macroscopic empty space is scale-invariant}} %MDPI: Please state whether italic format is necessary. If not, please consider removing it. Same for all instancees where this issue occures below.
 was further preferred  (Maeder 2017a), %MDPI: Reference citations should not appear in the abstract. Please check and revise.
 the resulting gauge is also supported by an action principle. Cosmological equations and a modified Newton equation were then derived. In short,  except in  extremely  low density regions, the scale-invariant  effects are largely dominated by Newtonian effects. However, their cumulative effects may still play a significant role in cosmic evolution. The  theory contains no ``adjustment parameter''. In this work, we gather concrete observational evidence that scale-invariant effects are present and measurable in astronomical objects spanning a vast range of masses (0.5  M$_{\odot} <$ M $< 10^{14}$  M$_{\odot}$) and  an equally impressive range of spatial scales (0.01 pc $<$ r $<$ 1 Gpc). Scale invariance  accounts for  the observed excess in velocity in galaxy clusters with respect to the visible mass, the relatively flat/small slope of rotation curves in local galaxies, the observed steep rotation curves of high-redshift galaxies, and the excess of velocity in wide binary stars with separations above 3000 kau found in Gaia DR3. Last but not least, we investigate the effect of scale invariance on gravitational lensing. We show that scale invariance does not affect the geodesics of light rays as they pass in the vicinity of a massive galaxy. However, scale-invariant effects do change the inferred mass-to-light ratio of lens galaxies as compared to GR. As a result, the 
discrepancies seen in GR between the total lensing mass of galaxies and their stellar mass from photometry may be accounted for. This holds true both for lenses at high redshift like JWST-ER1 and at low redshift like in the SLACS sample. Of note is that none of the above observational tests require dark matter or any adjustable parameter to tweak the theory at any given mass or spatial scale.}


\begin{document}

%%%%%%%%%%%%%%%%%%%%%%%%%%%%%%%%%%%%%%%%%%
% For any questions, please contact the editorial office of the journal or %support@mdpi.com. For LaTeX-related questions please contact %latex@mdpi.com.%\endnote{This is an endnote.} % To use endnotes, please %un-comment \printendnotes below (before References). Only journal Laws %uses \footnote.

%%%%%%%%%%%%%%%%%%%%%%%%%%%%%%%%%%%%%%%%%%
\section{Introduction} 
\label{intro}

The scale-invariant vacuum (SIV) theory rests on the simple and  fundamental  idea of examining whether it is possible, and maybe needed, to extend the group of invariances subtending gravitation theory, as suggested by Dirac \cite{Dirac73}. Dirac emphasized the following: {\it{``It appears as one of the fundamental principles in Nature that the equations expressing basic laws should be invariant under the widest possible group of transformations''}}. Thus, is there any other fundamental invariance, in addition to the Galilean invariance of Newton's Mechanics, the Lorentz invariance of special relativity, and the general covariance of general relativity (GR)? Clearly, if another such invariance exists,  its effects  must be vanishingly  small in current life as well as in the dynamics of the solar system, and might only apply  in rather restrictive and particular conditions.

The initial developments of theory were performed by Dirac  \cite{Dirac73} and by Canuto et al. \cite{Canuto77},  who proposed a general  scale-invariant field equation. Scale invariance means that the basic equations do not change upon a transformation of the line element of the following form:
\begin{equation}
ds'\,=\,\lambda(x^{\mu})\,ds\, ,
\label{ds}
\end{equation}
where $\lambda(x^{\mu})$ is the scale factor, where $ds'$ refers to GR, and where $ds$ refers to the scale-invariant space.  
The properties of the theory  have been presented in \cite{Maeder17a,MaedGueor23} (see also Section \ref{th}). As to the observational tests, several have been successfully studied, with interesting results \cite{MaedGueor20a,GueorMaed24}. Scale-invariant models \cite{Maeder17a} do predict an accelerated cosmological expansion, not far but still different from $\Lambda$CDM models, with other additional and positive cosmological tests \cite{Maeder17a}. Noticeably,  the growth of density fluctuations after recombination occurs quite rapidly without the need for dark matter \cite{MaedGueor19}. This may account for the occurrence of large galaxies at high redshift, as shown by recent JWST observations. The new equation of motion \cite{MBouvier79,MaedGueor23} in scale invariance predicts  a  tiny  additional acceleration that applies in the velocity direction:  it favors  accelerated expansion for outward motions, and collapses in a contraction.

Section \ref{th} presents the historical sequence of developments of the scale-invariant theory, as well as its most important properties and predictions that are relevant to our tests. Section \ref{clusters} focuses on the dynamics of galaxy clusters and Section \ref{rotgal} explores the rotation curves of galaxies, both at low and high redshifts. The velocity excesses recently found for wide binary stars in Gaia DR3  are examined in Section \ref{wide}.  Section \ref{lensing} analyses the differences seen between the photometric and lensing mass determinations of galaxies. Finally, Section \ref{concl} gives the conclusions.

%%%%%%%%%%%%%%%%%%%%%%%%%%%%%%%%%%%%%%%%%%
\section{The Origin and Properties of the Scale-Invariant Theory}  
\label{th} 

Scale invariance first appeared with Galileo Galilei, who discovered that {\emph{the natural physical laws are not scale-invariant}}, hence pointing to the existence of scaling laws in physics, e.g., \cite{Peterson02}.  Indeed, the presence of matter in a system, by fixing some scales, tends to suppress scale invariance, as pointed out  by Feynman \cite{Feynman63}. Scale invariance was  considered  by Weyl \cite{Weyl23} and Eddington \cite{Eddington23} in an attempt to account for electromagnetism by a geometrical property of space--time.  Their  original proposal was then abandoned because Einstein \cite{Einstein18} had shown that, in this case, the properties of a particle would  depend on its past world line, so that  atoms  in an electromagnetic field would not show sharp lines. Scale invariance has often reappeared in theoretical  physics, e.g., in Einstein--Cartan gravity with possible implications for fermionic dark matter and properties of the Higgs boson \cite{Karananas21}. The notion of scale invariance also appears in the study of fractal  distributions in physics and  astrophysics.  For example, in astrophysics, the initial density fluctuations  created by the inflation has been advocated for; this also appears in the distribution of voids and matter density in the Universe %Please check that intended meaning has been retained
\cite{Einasto89}.

The Weyl Integrable Geometry (WIG) was proposed by Dirac (1973) \cite{Dirac73} and by Canuto et al. (1977) \cite{Canuto77} (see also Bouvier and Maeder (1978) \cite{BouvierM78}). The length, $\ell$, of a vector transforms the same way as in Equation (\ref{ds}), and in the transport of a vector from a point $P_1(x^{\mu})$ to a nearby point  $P_2(x^{\mu}+dx^{\mu})$, the length $\ell$ of a vector is assumed to change by  $d\ell \, = \, \ell \, \kappa_{\nu} \, dx^{\nu}.$ The term $\kappa_{\nu}$ is called  the coefficient of metrical connection; it is a fundamental characteristic of the geometry, alike  to $g_{\mu \nu}$  (in GR $\kappa_{\nu}=0$). The  integrability condition in WIG makes it so that  Einstein's criticism does not apply%Please check that intended meaning has been retained
: the coefficient $\kappa_{\nu}$ must be a perfect differential,  $\partial_{\mu} \kappa^{\nu}=\partial_{\nu} \kappa^{\mu}$. Thus, the integral of the change in length on a closed loop is zero, implying that the integration does not depend on the path followed; see, e.g., \cite{Maeder23} for more details.

The field equation in the scale-invariant theory was   demonstrated by Canuto et al. \cite{Canuto77} in the line of results by Dirac \cite{Dirac73}. More recently, this field equation has been obtained from an action principle \cite{MaedGueor23}. In the same way as the field equation in GR requires a metric to determine the geometry of a system, a gauging condition is needed in the scale-invariant theory to fix the scale factor, $\lambda$.  Indeed, in this context, it is useful to recall the following statement by  Einstein \cite{Einstein49}: {\it{``…the existence of rigid standard rulers is  an assumption suggested by our approximate experience, assumption which is arbitrary in its principle''.}} Dirac \cite{Dirac73,Dirac74}, as well as Canuto et al. \cite{Canuto77}, used the so-called ``Large Number Hypothesis'' (LNH) to fix $\lambda$.  Another gauging condition was later  preferred by Maeder \cite{Maeder17a}: {\emph{the macroscopic empty space is scale-invariant}}; hence, the name of scale vacuum  invariant (SIV)  theory is often employed. This choice is justified, since the usual equation of the state for the vacuum $p_{\mathrm{vac}}=- \varrho_{\mathrm{vac}} c^2$ is precisely the relationship permitting $\varrho_{\mathrm{vac}}$ to remain constant for an adiabatic expansion or contraction \cite{Carroll92}. This condition leads  to some differential equations, the solution of which  gives  the scale factor  $\lambda=t_0/t$  \cite{Maeder17a}, where $t$ is the time of cosmological models; this time is  shown, for example, in  Equation (\ref{Jesus}). Subsequently, the search of a minimum of the action integral for small $\lambda$-variations was shown in \cite{MaedGueor23} to lead to the same differential equations and  scale factor as the above-mentioned gauging condition, which thus receives a strong basis.

General relativity (GR) and Maxwell equations are indeed scale-invariant in the empty space without charges and currents, an often ignored property.  A single atom in the Universe is likely to not be sufficient to kill scale invariance in an infinite Universe. Maeder and Gueorguiev \cite{MaedGueor21a} have discussed the problem of the physical connections in a scale-invariant Universe, as well as the occurrence of  inflation, with the conclusion that scale invariance is certainly forbidden in Universe models with a  mean density, $\varrho$, higher than the critical $\varrho_{\mathrm{c}}= 3 H^2_0 / (8 \pi G)$. This is in agreement with the conclusions of Galileo Galilei and Feynman that matter kills scale invariance, with the difference that   scale invariance is  an open possibility \cite{MaedGueor21a} for Universe models with a  mean density lower than the critical level; this conclusion is in agreement with the cosmological solutions below. 

The equations of the cosmological models  \cite{Maeder17a} resulting from the scale-invariant field equation with the FLWR metric
 and the mentioned gauging condition      lead  to \cite{Jesus18}
\begin{equation}
a(t) \, = \, \left[\frac{t^3 -\Omega_{\mathrm{m}}}{1 - \Omega_{\mathrm{m}}} \right]^{2/3}\, \quad \mathrm{with} \;\;
t_{\mathrm{in}}= \Omega^{1/3}_{\mathrm{m}}.
\label{Jesus}
\end{equation}

\begin{figure}[H]

\includegraphics[width=13cm]{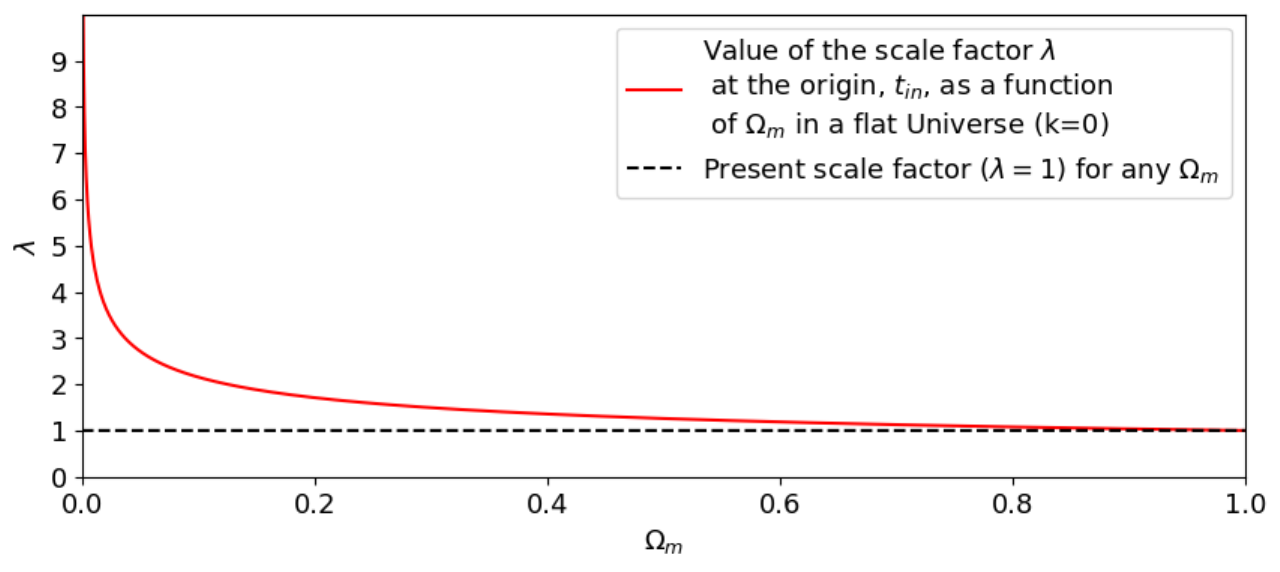}
\caption{The %MDPI: Figure was moved after its first citation. Please check and confirm.  Please add space before and after equal sign in figure. 
 red curve shows the scale factor $\lambda=t_{\mathrm{in}}/t_0$ at the origin  of the Universe ($a(t_{\mathrm{in}})=0$) for flat models, with $k=0$, as a function of the present-time $\Omega_{\mathrm{m}}$. The present scale factor at $t_0$ is $\lambda=1$ for any  $\Omega_{\mathrm{m}}$ (dashed black line). Thus, during the evolution of the Universe from the
Big Bang to present time, the value of $\lambda$ is only vertically moving, for a given $\Omega_{\mathrm{m}}$, from the red curve to the black dashed line.
This shows that, for increasing densities, the amplitudes of the variations of the scale factor $\lambda$ are very much reduced; $t_{\mathrm{in}}$ is given in Equation %MDPI: Information was revised from ``expr.'' into ``Equation''. Please check and confirm.
 (\ref{Jesus}).
}
\label{scale}
 
\end{figure}

\textls[-15]{The cosmological time $t$ varies between $t_{\mathrm{in}}= \Omega^{1/3}_{\mathrm{m}}$ at the origin, where $a(t_{\mathrm{in}})=0$, and the present time $t_0=1$, where $a(t)=1$. Equation (\ref{Jesus}) confirms  the absence of solution for $\Omega_{\mathrm{m}} \geq 1$, while it predicts  possible physical solutions for $\Omega_{\mathrm{m}} < 1$. Figure \ref{scale} shows that the SIV  effects  are drastically reduced for $\Omega_{\mathrm{m}}$ differing  from zero.  After a steep  decrease,  the  effects are smaller and smaller  for higher  $\Omega_{\mathrm{m}}$; they only totally vanish for $\Omega_{\mathrm{m}}  \geq  1$ (for a mean density  $ \geq 10^{-29}$ g  cm$^{-3}$). This is the scheme  we are proposing for the breakdown of scale invariance mentioned by Galileo Galilei and Feynman.  Since our Universe, according to all current results (e.g., Planck Collaboration \cite{Planck20}) has a mean density $\varrho$ such that  $(\varrho/\varrho_{\mathrm{c}})= \Omega_{\mathrm{m}} \leq 0.30$, it appears that {\emph{scale invariance is  a possibility}}. There is a relation between the dimensionless time scale $t$ and the current time units $\tau$ (in Gyr or seconds):}
\begin{equation}
\frac{\tau - \tau_{\mathrm{in}}}{\tau_0 - \tau_{\mathrm{in}}} = \frac{t - t_{\mathrm{in}}}{t_0 - t_{\mathrm{in}}},
\end{equation}
expressing that the age fraction with respect to the present age is the same in both timescales. This means that
\begin{equation} 
t \,= \,t_{\mathrm{in}} + \frac{\tau}{\tau_0} (1- t_{\mathrm{in}}).
\label{time}
\end{equation}

Times $t$ and $\tau$ are the same physical time: simply, when the first goes from $t_{\mathrm{in}}$ to 1, the second goes from 0 to 13.8 Gyr. 
The dynamical tests presented in this work are based on the equation of motion, derived from the geodesic for weak fields   \cite{Dirac73,Canuto77},  also derived from an action principle by \cite{BouvierM78}. This equation of motion, corresponding to a modified Newton equation, is---at the present time---$\tau_0$ \cite{MBouvier79,MaedGueor23}:
\begin{equation}
 \frac {d^2 \bf{r}}{d \tau^2}  = - \frac{G  M(\tau) }{r^2}  \frac{\bf{r}}{r}   + \frac{\psi_0}{\tau_0}   \frac{d\bf{r}}{d\tau} , 
 \quad \mathrm{with} \; \; \psi_0 =1-\Omega^{1/3}_{\mathrm{m}} \,,
\label{Nvec4}
\end{equation}
\noindent 
where $\tau_0$ is the present age of the Universe in current time units. The additional term on the right of Equation (\ref{Nvec4}) is an acceleration  in the direction of the motion, we call it {\it{the dynamical gravity}}; it is  usually very small, since $\tau_0$ is very large. This term, proportional to the velocity,  favors collapse during a contraction,  and  outwards acceleration in an expansion (in fact, a corresponding acceleration  term appears in the cosmological Equations \cite{Maeder17a}). The parameter  $\psi_0$ applies  at  the present epoch; for other times, $\tau$, instead of $\psi_0$, one has 
  \begin{equation}
\psi(\tau) = \frac{t_0-t_\mathrm{in}}{\left( t_\mathrm{in} +   \frac{\tau}{\tau_0} (t_0-t_\mathrm{in})\right) }.
\label{psit}
\end{equation} 

  The ratio of the new term to the Newtonian one behaves like  $\sqrt{\frac{\varrho_c}{\varrho}}$ in systems at equilibrium  \cite{Maeder17c}.  Thus, at the present epoch, the  dynamical gravity is generally negligible. However, cumulative effects  may appear, especially at low densities, on long time scales of the order of $H^{-1}_0$.

From the conservation laws \cite{Maeder17a}, it comes out that the potential  $\Phi=  GM/r$ is conserved over the evolution of the Universe, while the masses, alike in special relativity,  are variable,   scaling here like $M \sim \lambda^{-1}  \sim t$. These variations are very limited, due to the small  range of $\lambda$  (Figure \ref{scale}). As an example, there is a change in mass by a fraction of 0.0025 or 0.0046 over the last hundred million years for $\Omega_{\mathrm{m}}= 0.30$ and 0.05, respectively (new solar models %MDPI:Footnotes should not be used. Information was added within main text in parethesis. Please check and confirm.
 by Prof. G. Meynet show that the effects on the present solar luminosity are negligible, leading to solar models deviating by less than 1\% from the observed value; see also Section 3 %MDPI: Please cofirm that Section 3 does not refer to Section 3 of this manucript, thus not needing to be likend.
 of \cite{Maeder77}, which showed that the small changes in the product $GM \sim t$ surprisingly have no significant effect on the present solar luminosity). The above considerations are theoretical ones, suggesting that some effects of scale invariance are possible in our low-density Universe. The answer belongs to experiments. This is why here we are studying  several observational tests, that were not considered in \cite{GueorMaed24}.

The mathematical tools of the scale-invariant theory may look  a bit more complex than those in GR, since  the mathematical developments of  the scale-invariant theory require, instead of tensors, the introduction of cotensors, i.e., mathematical tools which, in addition to the general covariance of GR, also include the scale covariance \cite{Dirac73, Canuto77, BouvierM78} (for a brief summary see also \cite{Maeder23,Canuto77}).  This  slightly modifies the Newton Equation (\ref{Nvec4}), as shown nearly half a century ago \cite{MBouvier79}
 (see also 
\cite{Maeder17c,MaedGueor23},    some conservation laws in \cite{Canuto77,Maeder17a}, as well as the cosmological Equations in \cite{Maeder17a}). Despite these differences,  the scale invariance theory is  in full agreement with a fundamental statement about gravitation  by Einstein \cite{Einstein49}, who considered  that {\emph{a simple fundamental mathematical principle must  determine the equations}}.

There are interesting similarities and differences between  the scale-invariant theory and the current
scalar-tensor theories. The similarity is
that the scale-invariant theory effectively contains a scalar field $\varphi$, in addition to the tensor formalism.
In general, in scalar-tensor theories of gravitation \cite{Fujii03,Clifton12}, the scalar field is defined independently, in order to specifically study some   particular physical situations, such as  the neighborhood of black holes or inflation. The specific difference of the scale-invariant theory is that the definition of the scalar field $\varphi$ is based on the scale factor $\lambda$,
\begin{equation}
 \varphi  =  \ln \lambda, \quad \quad \mathrm{with} \; \;  \kappa_{\nu} = - \frac{\partial \ln \lambda}{\partial x^{\nu} }.
\end{equation}

The properties of $\lambda$ are determined by a gauging
condition. The choice we made (see above) is  resting on the properties of the macroscopic vacuum and is therefore a universal function 
$\lambda=1/t$, independent on the matter properties.  Thus, 
since the macroscopic vacuum is assumed the same everywhere, so does
$\lambda$ and  thus $\varphi$.
Thus, there is no freedom to choose a particular field for a specific
purpose, and this is a constraining property of the scale-invariant theory, which is thus different from usual scalar-tensor theories.
We note that 
this field $\varphi$ or $\lambda$ is acting in the Newton-like approximation, as well it may  may play the role of the rolling
field in the inflation  \cite{MaedGueor21a}. 
Considerable theoretical developments of scalar-tensor theories
have recently have been performed in terms of Lorentz fiber-bundles by Ikeda et al. \cite{Ikeda19}.
The fiber-bundle locally represents a multidimensional  space as a projection on another globally different topological structure.
 This  framework
is providing a new valuable approach of the 
cosmological dynamics for the scalar-tensor theories. It has been applied
by \cite{Ikeda19} both to the inflation and to the  later stages of cosmological evolution, where it introduces some  extra terms in the cosmological equations. The study of this new possible 
theoretical  approach  in the scale-invariant context is however
beyond the scope of the present paper. At this stage, we note that
that the scale-invariant theory is, like GR, invariant to Lorentz
transformations, since the change in scale applies the same to the
time and spatial length, keeping the velocities constant.\\

%%%%%%%%%%%%%%%%%%%%%%%%%%%
\section{The Mass Excess in Clusters of Galaxies}  
\label{clusters}
%%%%%%%%%%%%%%%%%%%%%%%%%%%

For the derivations  and  information about the equations used in this work, we refer the  reader to  \cite{Maeder17a,MaedGueor23}.	 The possibility of dark matter appeared in 1933 with the work of Zwicky \cite{Zwicky33}. However, it was not until  around 1980 that the hypothesis of dark matter was generally recognized%Please check that intended meaning has been retained
. Indeed, large excesses of Virial masses in clusters have repeatedly been found for nearly a century \cite{Zwicky33, Karachentsev66, Bahcall74, Blindert04, Sohn17}. Virial masses in excess with respect to stellar luminosities have been confirmed, reaching a factor of about 30--50  \cite{Proctor15} for large clusters with masses between $10^{14}$ and $10^{15}$ M$_{\odot}$. This implies that the stellar mass fraction is only of the order of 0.02--0.03. Today, it appears that dark matter has been progressively upgraded from its initial status of an hypothesis to that of a material reality.

X-ray observations of the emitting gas  allow estimates of  the gas mass  fraction (typically 0.10--0.15) in clusters \cite{Lin12, Leauthaud12, Gonzalez13, Shan15, Ge16, Chiu16}, which  largely dominates over the stellar mass fraction mentioned above.  These two fractions contribute to the baryon mass fraction, $f_{\mathrm{bar}}$,  which is typically around 0.15--0.18, a factor of about  6 smaller than  unity; this result is in agreement with the Planck Collaboration \cite{Planck20} for matter, baryon, and dark   matter density parameters: $\Omega_{\mathrm{m}}=0.315 \pm 0.007$, $\Omega_{\mathrm{b}}=0.050 \pm 0.0002$, and  $\Omega_{\mathrm{DM}}=0.265 \pm 0.002$. Noticeably, a relation between the total dynamical mass of clusters and  the baryonic masses of central regions  has also been found  \cite{Chan22}. 
 
We may attempt a numerical estimate of the relation between  masses and velocity dispersions in the scale-invariant theory, using the fact that the scale factor, and the quantities which depend on it, have limited variations and may be considered as adiabatic invariants. As the Virial theorem does not apply in an expanding system,  we start directly from the equation of motion (\ref{Nvec4}) and consider a galaxy $i$  with a velocity $\upsilon_i$ that is attracted  by another $j$-one at time $\tau$ and distance $r_{ij}$. Its acceleration at any time $\tau$ occurs according to Equation (\ref{Nvec4}); further,
 \begin{eqnarray}
 \frac{d\upsilon_i}{d \tau} = \frac{G  m_j(\tau)}{r^2_{ij}}    + \frac{\psi(\tau)}{\tau_0}   \upsilon_i ,  \quad \mathrm{with} \; \;
\psi(\tau)  = \frac{1}{\frac{\Omega^{1/3}_{\mathrm{m}}}{1-\Omega^{1/3}_{\mathrm{m}}}+ \frac{\tau}{\tau_0} } .  \label{Nv1}\\
\mathrm{thus} \quad \frac{\psi(\tau)}{\tau_0} = \frac{1}{\tau_{\Omega}+\tau}\, , \quad \mathrm{with} \; \; 
\tau_{\Omega}=\tau _0 \,\frac{\Omega^{1/3}_{\mathrm{m}}}{1-\Omega^{1/3}_{\mathrm{m}}}.
 \label{Nvecv}
\end{eqnarray}

The first equation is then multiplied by $\upsilon_i = dr_{ij}/d\tau$, integrated and summed up  on all the other $(N-1)$  $m_j$-masses. As the variations of  mass $m_j(\tau)$ are  very limited, we consider a mean value  $ p \, m_j(\tau_0)$ of  masses $m_j$  at half the age of the Universe; thus, $p$ is given by $p \approx (1+\Omega^{1/3}_{\mathrm{m}})/2$.  One  also has, when summing all the N  $i$-masses,
\begin{equation}
\frac{1}{2}\sum_i \upsilon^2_i  = \frac{1}{2}  p \sum_i \sum_{j\neq i} \frac{G\,  m_j (\tau_0)} {r_{ij}} + 
\sum_i\int  \frac{1}{\left[\tau_{\Omega}+
\tau \right]} \, \upsilon^2_{i} d\tau \, .
\label{en1}
\end{equation} 

The factor $1/2$ in front of the double summation is necessary  in order to   not account for the same masses twice%Please check that intended meaning has been retained
. We divide the above equation by $N$. The   term on the left becomes $\frac{1}{2} \overline{\upsilon^2}$.  The first term on the right gives the potential energy $ p q G M/R$ of the cluster with a total present mass   $M$ and a radius $R$, with $q$ as a structural factor, depending on the density profile ($q=3/5$ for an homogeneous distribution). The last term is an adiabatic invariant, for which we take an average over the lifetime of the cluster between $\tau_1$ and $\tau_0$, where $\tau_1$ is the formation epoch. We integrate it as follows:
\begin{eqnarray}
  f \overline{\upsilon^2}\, \int ^{\tau_0}_{\tau_{1}}  \frac{ d\tau}{\left[\tau_{\Omega}+\tau\right]}\,= 
  f\overline{ \upsilon^2} \,\ln  \frac{\tau_{\Omega}+\tau_0}{\tau_{\Omega}+\tau_1} = 
  f \overline{\upsilon^2}  \times  x(\Omega_{\mathrm{m}}), \,
   \mathrm{with} \; \; x(\Omega_{\mathrm{m}})= 
  \ln  \frac{\tau_{\Omega}+\tau_0}{\tau_{\Omega}+\tau_1}. \quad  \quad
 \label{t3}
  \end {eqnarray}

The   factor $f$ accounts  for the integration of $\upsilon^2$ over time $\tau$; a value $f=0.5$ is likely a minimum value. The factors $x(\Omega_{\mathrm{m}})$ are given by Equation (\ref{t3}), taking a ratio $\tau_1/\tau_0= 3\%$, which corresponds to  galaxy formation at a cosmic time of 400 Myr, with the choice of this  initial time not being critical. The values of $ x(\Omega_{\mathrm{m}})$ are $ x(\Omega_{\mathrm{m}}) =$ 0.387, 0.516, 0.735, and 0.950, respectively, for $\Omega_{\mathrm{m}}=$ 0.30, 0.20, 0.10, and 0.05. The integration constant on the first term in Equation (\ref{en1}) may be considered as zero, since a zero velocity  is an appropriate value for the end of the collapse. The same is true for the potential, which vanishes at large distances. Equation (\ref{en1})  can be written as follows:
\begin{eqnarray}
\overline{\upsilon^2} \, = \,  2 \,\frac{ pq \,G M_{\mathrm{SIV}}}{R}+
 2  \, x(\Omega_{\mathrm{m}}) \times f \overline{\upsilon^2} \,,
%\mathrm{or} \quad  \overline{ \upsilon^2} \, =\frac{2 \, {pq \,G M(\tau_0)}/{R}}{1 - 2 f x(\Omega_{\mathrm{m}})}. \quad
\label{en22}
\end{eqnarray}
where $M_{\mathrm{SIV}}$ is the present mass in the scale-invariant theory. In  the standard case,  a value $ M_{\mathrm{std}}$ would be derived from the same value of the velocity dispersion  $\overline{\upsilon^2} $  by  the usual relation, $\overline{\upsilon^2} \, = \,  2 \, q \,G M_{\mathrm{std}}/{R}$. From Equation (\ref{en22}), $M_{\mathrm{SIV}}$ is evidently smaller than $M_{\mathrm{std}}$. The ratio of the two  mass estimates is given by
\begin{equation}
\frac{M_{\mathrm{std}}}{M_{SIV}} \,  \approx \, \frac{p}{ \left[1 - 2 \, f \,x(\Omega_{\mathrm{m}}) \right]},
\label{mratio}
\end{equation}
all masses being considered at the same time for not too distant clusters.
The same structural factor $q$ is taken in the two cases. The statistical factors,  which account for the fact that the observed velocities are the radial ones, are the same in the two theories. The factor $p$ has no effect in the term responsible for the divergence. One has\linebreak  $p$ = 0.835, 0.792, 0.732, and 0.684  for  $\Omega_{\mathrm{m}}=$ 0.30, 0.20, 0.10, and 0.05, respectively. 

The above expression shows that  the standard mass estimates are generally much larger than the SIV ones. This is a result of the cumulative contribution over the ages of the additional acceleration  in the equation motion.  This may explain why  ratios $\frac{M_{\mathrm{std}}}{M_{SIV}}$ as high as 80   \citep{Karachentsev66} or  mass--luminosity ratios $M/L$ of about 700 \citep{Bahcall74} have been obtained in the past. The ratio  $\frac{M_{\mathrm{std}}}{M_{SIV}}$ is  particularly large and often diverging for low values of  $\Omega_{\mathrm{m}}$.  For  $\Omega_{\mathrm{m}}=0.05$ and $f=0.5$, the ratio is $\frac{M_{\mathrm{std}}}{M_{SIV}}=13.7$. For higher $f$-values, it diverges, since the denominator of (\ref{mratio}) vanishes.  For larger $\Omega_{\mathrm{m}}$, e.g.,  0.20, the mass ratios are 4.5, 11.1, and $\infty$ for $f=$0.8, 0.9, and 1.0. Thus,  if scale-invariant effects are present in nature, we  expect that applying standard mass estimates result in large overestimates of the  masses of galaxy clusters. 

Within the scale-invariant theory, there is no need to advocate for dark matter to account for the large observed velocity dispersions in  galaxy clusters. This result, joined  to the other ones noted in
the introduction  and to those further presented in this article, offers direct support  to the scale-invariant theory. 
Numerical simulations would be very helpful to better quantify the various effects already suggested by the above analytical approach.

%%%%%%%%%%%%%%%%%%%%%%%%
\section{Galactic Rotation}  
\label{rotgal}
%%%%%%%%%%%%%%%%%%%%%%%%
\subsection{Binet Equation, Angular Momentum, Secular Variations, and Velocities}   \label{Sinet}
	
The concept of dark matter has been greatly promoted \cite{Trimble87} by the observations of the flat, or relatively flat, rotation curves of galaxies over a few tens of kpc (see review by  \cite{Sofue01}). The observations much differ from the standard predictions, which---in the outer regions of spiral galaxies---predict  orbital velocities $\upsilon$ that behave similarly to  Keplerian in $1/\sqrt{r}$, where $r$ is the galactocentric radius. This was a severe disagreement 
with respect to the predictions  of classical mechanics.

This is why it is worth exploring the potentialities of the scale-invariant theory in this context. 
The properties of galactic rotation in the outer layers are mainly determined by those of a  two-body system with  a central mass and a test particle. The dynamic of wide binary stars in Section \ref{wideobs} is also governed by the same equations; thus, it is worth briefly  considering the problem in detail. These properties were originally derived nearly half a century ago \cite{MBouvier79} in the system of the cosmological dimensionless $t$-units. Here, we give a demonstration of this in the current $\tau$ units (years and seconds), which brings some changes in the formal writing of equations. From the equation of motion  in current units (Equation \ref{Nvec4}),  the corresponding two equations in plane polar  coordinates $(r,\vartheta)$ are
\begin{equation}
\ddot{r}-r\dot{\vartheta}^2   =  -\frac{GM}{r^2}+\frac{\psi}{\tau_0} \frac{dr}{d\tau},  \quad \quad \mathrm{and} \quad
r \ddot{\vartheta}+ 2 \dot{r} \dot{\vartheta}  =  \frac{\psi}{\tau_0} r \dot{\vartheta}.
\label{2pol}
\end{equation}

When not specified, $\psi$ is $\psi(\tau)$, as given by Equation (\ref{Nv1}); similarly, $M$ is $M(\tau)$. The 2nd of the above equations has the following for an integral:
\begin{equation}
\frac{\psi}{\tau_0} \,r^2 \dot{\vartheta} = \mathcal{L},
\label{consang}
\end{equation}
$\mathcal{L}$ is a constant  in $[{\rm cm^2 \, rad \, s^{-2}}]$.  This  is a generalization of the classical angular momentum conservation $r^2 \dot{\vartheta} = {\rm const.}$, by mass unit. It shows that $r^2 \dot{\vartheta}$ slowly increases with time. We now express the content of the three above equations into one single relation.  With  Equation (\ref{consang}), the derivative of $r$ with respect to $\vartheta$ can be written as
\begin{equation}
r' \equiv \frac{dr}{d \vartheta} = \frac{dr}{d\tau} \, \frac{d\tau}{d \vartheta}= \frac{\dot{r}\; r^2 \psi}{\mathcal{L} \; \tau_0} \quad
\mathrm{thus} \; \; \dot{r}=\frac{r' \mathcal{L} \, \tau_0}{r^2 \,   \psi}\,.
\label{dp}
\end{equation}

This kind of relation between $r'$ and $\dot{r}$ can be extended to $\ddot{r}$,
\begin{equation}
\ddot{r}= \frac{\mathcal{L}^2 \,r'' \tau^2_0}{r^4 \, \psi^2} - \frac{2 r'^2 \mathcal{L}^2 \tau^2_0}{r^5 \psi^2}+
\frac{r' \mathcal{L}}{r^2}\,,
\label{dd}
\end{equation}
since $\dot{\psi} =-\psi^2/\tau_0$. Expressing the first equation  of (\ref{2pol}) with (\ref{dd}), we obtain,
\begin{equation}
\frac{\mathcal{L}^2 \, \tau^2_0}{r^4 \, \psi^2}\left(r''-2 \frac{r'^2}{r}\right) -\frac{\mathcal{L}^2 \, \tau^2_0}{r^3 \, \psi^2}+
 \frac{G \, M}{r^2}=0 \,,
\label{r2}
\end{equation}
where---remarkably and fortunately---the last  terms in Equations (\ref{Nvec4}) and (\ref{dd}) have simplified. Now, writing this last equation with $u=1/r$ is leading simply to,
\begin{equation}
u'' + u = \frac{GM\, \psi^2}{\mathcal{L}^2 \, \tau^2_0}\,.
\label{Binet}
\end{equation}

This is the modified form of the well-known Binet equation, describing planetary trajectories.  We will again find it in the study of lensing as Equation (\ref{compl}), also containing both post-Newtonian 
and scale-invariant terms.   If the 2nd member is equal to $0$, the solution is $u= C \cos \vartheta$, the polar equation of a straight line. Now, we search a solution of the form $u= p+C \cos\vartheta$ and obtain  $p=  \frac{GM\, \psi^2}{\mathcal{L}^2 \, \tau^2_0}$ $[{\rm cm}^{-1}]$, known as  the {\emph{latus rectum}} in  the  geometrical construction of conics. $C$ is also a constant in $[{\rm cm}^{-1}]$. Thus, the solution $r(\vartheta)$ is
\begin{equation}
r=\frac{1}{\frac{GM\, \psi^2}{\mathcal{L}^2 \, \tau^2_0}+C \cos \vartheta}\, \quad \quad \mathrm{or} \quad
r= \frac{r_0}{1+e \cos \vartheta}  , 
\label{traj}
\end{equation}
where $e$ is the eccentricity, determined by the initial conditions. It is dimensionless and therefore a scale-invariant quantity.
If the eccentricity is zero, then $r_0=1/p$, 
\begin{equation}
\; e= \frac{\,C \mathcal{L}^2 \, \tau^2_0}{GM\, \psi^2}\,,\quad \quad \mathrm{with}  \; \;
r_0  \, = \, \frac{\mathcal{L}^2 \, \tau^2_0}{GM\, \psi^2}\,,
\label{r0}
\end{equation}
$r_0$ in $[{\rm cm}]$ is the radius  of the  circular orbit. It  behaves like $1/\lambda$, thus  increasing with time. For elliptical solutions, with $e<1$, the semi-major and semi-minor axes, respectively, are $a=\frac{r_0}{(1-e^2)}$ and $b=\frac{r_0}{\sqrt{1-e^2}}$. These  are not classical conics with constant parameters (except for $e$). As the result of the time dependence of $r_0$, they  are {\emph{expanding conics}}  \cite{MBouvier79}. From (\ref{r0}), one has, considering a distance $r$ which may apply to a conic or a circle,
$ \frac{\dot{r}}{r}= -\frac{\dot{M}}{M}-2 \frac{\dot{\psi}}{\psi}\,.$
Since $M \sim  t$, and given Equation \ref{time}, one has
\begin{equation}
M(\tau)= M(\tau_0) \, [t_{\mathrm{in}}+\frac{\tau}{\tau_0}(1-t_{\mathrm{in}})], \quad 
\quad \frac{\dot{M}}{M}= \frac{\psi}{\tau_0}\, , \quad \mathrm{and} \; \;
\frac{\dot{r}}{r}=  \frac{\psi}{\tau_0} \,,
\label{drdt}
\end{equation}
with $\dot{\psi} =-\psi^2/\tau_0$. This implies that the distance $r$, or the semi-major axis $a(\tau)$ of the elliptical motions, behaves like $r(\tau)= r(\tau_0)\,  [t_{\mathrm{in}}+\frac{\tau}{\tau_0}(1-t_{\mathrm{in}})].$ The semi-major axis is therefore increasing linearly in time $\tau$, consistent with the variation of $a(t)=a(t_0) \times t $ obtained in the $t$-scale \cite{MBouvier79}. 

We now turn to the orbital period $T$, which, for simplicity, is considered in the circular case. With the angular velocity $\dot{\vartheta}= 2 \pi/T$ (for an elliptical orbit $\dot{\vartheta}$ would be the mean angular velocity)  and the angular momentum conservation $\dot{\vartheta}=\frac{\mathcal{L} \tau_0}{r^2 \psi}$, one is led to
\begin{equation}
\frac{\dot{\vartheta}}{\vartheta}= -2 \frac{\dot{r}}{r} -\frac{\dot{\psi}}{\psi}=
-2  \frac{\psi}{\tau_0} +  \frac{\psi}{\tau_0}= - \frac{\psi}{\tau_0},\quad \quad  \quad \mathrm{thus} \; \;\frac{\dot{T}}{T}= 
-\frac{\dot{\vartheta}}{\vartheta}= \frac{\psi}{\tau_0}.
\label{TT}
\end{equation}

This implies that the period $T$ is varying, like $T(\tau)= T(\tau_0)\,  [t_{\mathrm{in}}+\frac{\tau}{\tau_0}(1-t_{\mathrm{in}})]$. Thus, the mass $M$,  the parameters $a$, $b$, and $r_0$, and the orbital period $T$ all have the same kind of time dependence.

Let us now turn to the orbital velocity. Starting from (\ref{consang}), for the orbital velocity, we have  $\upsilon=r\dot{\vartheta}= \frac{\mathcal{L} \,\tau_0}{\psi \, r}$. Again considering a circular orbit, by introducing the expression of $r_0$ given in Equation (\ref{traj}), one obtains
\begin{equation}
\upsilon = r_0 \dot{\vartheta} = \frac{\mathcal{L} \,\tau_0 \,GM \psi^2 }{\psi \,  \mathcal{L}^2 \,\tau^2_0 }= \frac{ \ GM \psi}{  \mathcal{L} \tau_0 },
\quad \mathrm{and} \; \; \upsilon^2=  \frac{ \ G^2 M^2 \psi ^2}{  \mathcal{L}^2 \tau^2_0 }= \frac{GM}{r_0}.
\label{vps}
\end{equation}

The time dependencies  of  $M$ and $\psi$  cancel each other out; we have  used (\ref{traj}). The orbital velocity has the usual  standard expression; it is  scale-invariant, keeping constant in time during  secular orbital expansion.  This is  consistent with the fact that the potential is a conserved quantity in the theory. However, it does not imply that the velocity dispersion is the same, as seen in Section \ref{clusters}. Regarding  Equation (\ref{vps}), it is interesting to note  that  there is in fact a very subtle interplay in the scale-invariant theory:  the tangential  (additional) dynamical acceleration   $\frac{\psi_0}{\tau_0}   \frac{d\bf{r}}{d\tau}$  exactly compensates for the usual slowing down due to the orbital expansion so as to keep the orbital velocity constant during the secular  expansion.

\subsection{Recent Observations}

The bulk of observations of spiral  galaxies show flat rotation curves up to galactic distances of a few tens of kpc according to Sofue and Rubin \cite{Sofue01}. The results by Huang et al. \cite{Huang16} show an extended  rotation curve  up to about 100 kpc  based on the observations of about 16,000 %MDPI: ` was revised into comma as in numbers with more than 4 digits. Please check and confirm. Same for all instances below.
 red clump giants in the outer disk and 5700 K giants in the halo  (Figure \ref{MWcompl}). A proper account of the anisotropic motions is performed in the analysis, which also permits one to see the undulations due to spiral arms in the inner region; but, on the whole, this part could be regarded as globally flat. Beyond 26 kpc a progressive decrease is observed up to  distances as large as 100 kpc. 

More recent results by Eilers et al. \cite{Eilers19} and by Jiao et al. \cite{Jiao23} find steeper curves for the 5 to 25 kpc inner distance interval. The study  by Eilers is based on the 6D data (location and velocities) of about 23,000 red giant stars where the velocity curve is   determined from the Jeans equation for an axisymmetric gravitational potential. These authors found a decreasing velocity of 1.7 $(\pm  0.1)$ km/s per kpc. They estimated   a value of 7.25 ($\pm 0.26)\times 10^{11}$ M$_{\odot}$ for the total mass of the Milky Way, and found that  dark matter is dominating for radii larger than about 14 kpc. The results by \cite{Jiao23} %MDPI: Please provide reference citation after year (e.g.: [1], [2,3], et.).
are based on Gaia DR3. They find  a steeper decrease in the velocity  of 30 km s$^{-1}$ between 19.5 and 26.5 kpc, leading to a much smaller mass value of 2.06$^{(+0.24)}_{(-0.13)} \times 10^{11}$ M$_{\odot}$.  Up to 20 kpc from the center, the data by both groups agree with the lowest points of the wiggles due to galactic arms by Huang et al., while the wiggles are absent from the other two axisymmetric  studies  (Figure \ref{MWcompl}).
	
\subsection{Predicted Effects in the Galactic Evolution}

The above secular variations  and velocity constancy allow us to reconstruct the  evolution of 
galaxies, for the part due to scale invariance \cite{Maeder17c, Maeder24}. In particular, we consider  the two following effects:
	\begin{itemize}
	\item The orbital  radius 
	(here the galactocentric radius)  increases with time  (Equation (\ref{drdt})). 
	\item The orbital velocity $\upsilon$ of a given star 
	around the galactic center remains constant during the additional expansion  (Equation (\ref{vps})). 
	\end{itemize} 
	
The two above effects contribute to shape the rotation curves of galaxies starting from the initial velocity profile $\upsilon(r)$. From the initial profile, likely not far from  a steep   Keplerian shape,  the above effects  are progressively producing a flattening of the curve,   which increases with  the age of the galaxy and  also depends on the  other internal dynamical effects. A clear prediction is that {\emph{the steepness of the profile should increase with redshifts  $z$}}.

The time intervals considered in the evolution of galaxies are not negligible with respect
to the age of the Universe; thus, we cannot simply  write 
\begin{equation}
\frac{\Delta r}{\Delta \tau}  \, = \psi_0 \frac{r(\tau_0)}{\tau_0},\,
\end{equation}
as would be performed for small changes at the present time.  Instead, we must integrate Equation (\ref{drdt}) over time $\tau$ between $\tau_1$ (in Gyr) the epoch considered in the past with size $r_1$, and the present time $\tau_0$ with size $r_{\tau_0}$,
\textls[-25]{\begin{eqnarray}
	\int^{r_0}_{r_1} \frac{dr}{r} = \frac{(1-\Omega_ {\mathrm{m}})^{1/3}}{\tau_0} \int^{\tau_0}_{\tau_1} 
	\frac{d\tau}{\Omega_{\mathrm{m}}^{1/3}+\frac{\tau}{\tau_0}(1-\Omega_ {\mathrm{m}}^{1/3})}=  
	\int^{\tau_0}_{\tau_1} \frac{d\tau}{\tau_0  \frac{\Omega_ {\mathrm{m}}^{1/3}}{(1- \Omega_ {\mathrm{m}})^{1/3}}+ \tau}=
	\int^{\tau_0}_{\tau_1}\frac{d\tau}{\tau_{\Omega}+\tau},
\end{eqnarray}}
where we use the constant $\tau_{\Omega}$ defined in Equation (\ref{Nvecv}). The integration gives
\begin{equation}
	 \frac{r_0}{r_1} \,=\,  \frac{\tau_{\Omega}+\tau_0}{ \tau_{\Omega}+\tau_1}.
	\label{ratiogal}
\end{equation}

\begin{figure}[H]
 
\includegraphics[width=0.8\textwidth]{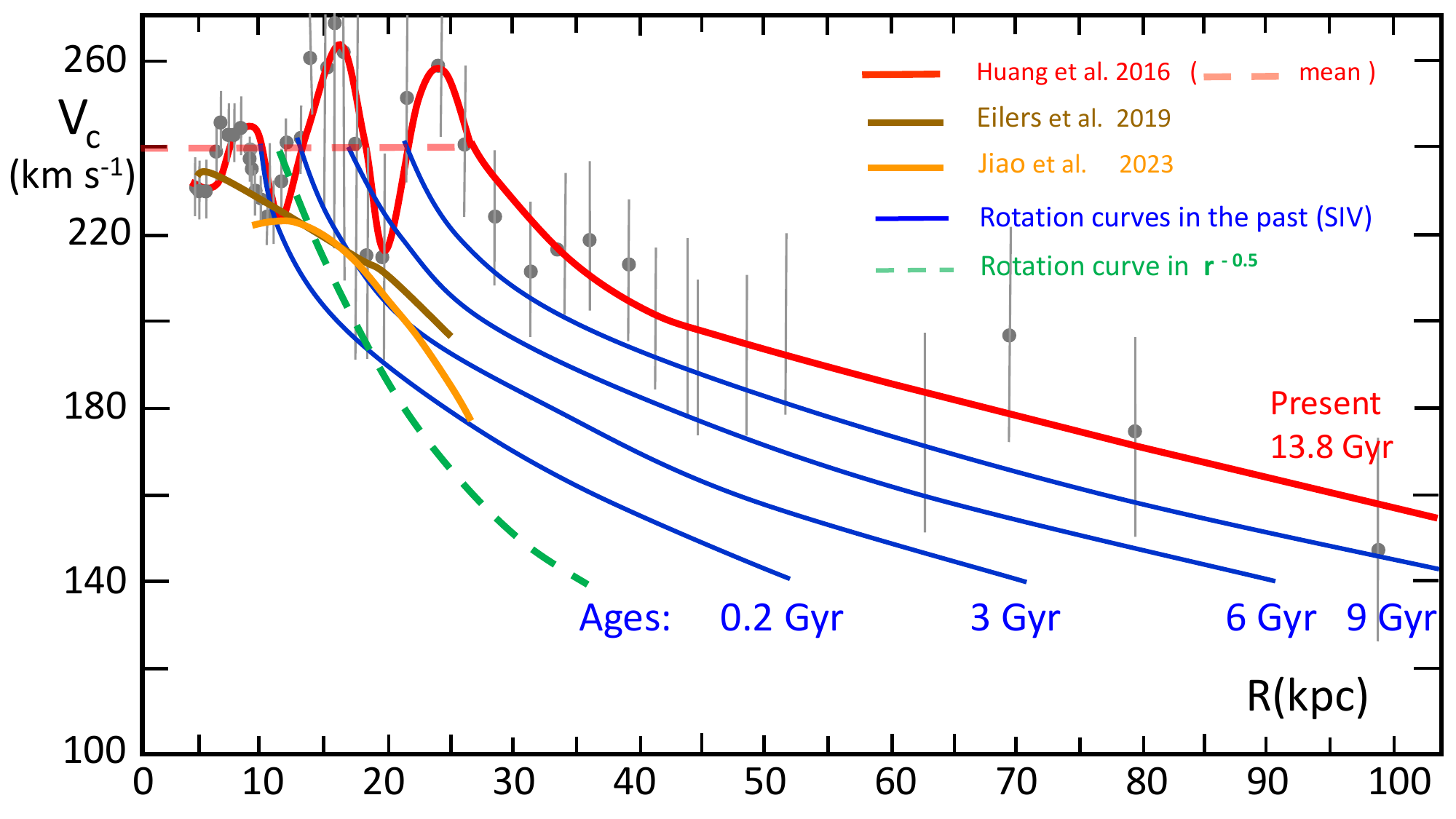}
\caption{Evolution %MDPI: Figure was moved after its first citation. Please check and confirm. Please change the hyphen (-) into minus sign ($-$, "U+2212"). e.g., "-1" should be "$-$1".
 of the rotation curve of the Milky Way. The gray points are the velocity
averages observed by Huang et al. \cite{Huang16}, with their error bars.  The thick red line represents a mean rotation curve. The thin dashed red line describes the flat mean of the wiggles of the  velocity distribution up to a radius of 26 kpc. The  brown and orange lines show, respectively, the results of Eilers et al. \cite{Eilers19} and by Jiao et al. \cite{Jiao23} for the inner galaxy. The blue  lines show the  rotation curves  predicted by  the scale-invariant theory for  different ages in the past history of the Universe, starting backwards from the red curve. Calculations have been performed with no dark matter in a  model with  $\Omega_{\mathrm{m}}=  \Omega_{\mathrm{b}}$, where $\Omega_{\mathrm{b}}=0.045$. The thick dashed green line shows a Keplerian curve  in  $1/\sqrt{r}$ near  the  age of galaxy formation.}
\label{MWcompl}
\end{figure}

For example, for  $\Omega_ {\mathrm{m}}=0.045 $ \cite{Planck20}, $\tau_0=13.8$ Gyr  and   an age  $\tau_1$ of 6 Gyr, one has $\tau_{\Omega}= 7.62$ Gyr and a scaling of the distances $\frac{r_0}{r_1}= 1.57$. This means that, at an age of 6 Gyr, the same velocity  as one on the red curve   of (Figure \ref{MWcompl}) was reached at a galactocentric distance $r_1$, which is a factor 1.57 smaller than the present  one. As an  illustration of  the predicted effects, we take the rotation curve given by Huang (2016) \cite{Huang16} (red curve in Figure \ref{MWcompl}). The above two rules lead to  the construction of the past rotation curves of galaxies at different past ages of 0.2, 3, 6, and 9 Gyr, as shown by the blue curves in this figure. We notice that they become steeper for further in the past. Among the blue curves, the left one corresponds to the early age of galaxy formation (0.2 Gyr). When  more detailed curves may come, the above process may also be applied and used for comparisons with the curves in high-redshift galaxies (see below where some observable trends are already  noticed). These results illustrate the cosmic evolution resulting from scale-invariant effects.  

The rather steep---but short---curves in the inner part, found more recently  by Eilers et al. \cite{Eilers19} and by Jiao et al. \cite{Jiao23}, could also be treated backwards in the same way as for Huang  data. They would lead to a very small core with a steep velocity gradient in the early evolutionary stages. We also note that these recent curves rather correspond to the minimum velocity of the dips associated with spiral arms. At high redshift, we  may expect the following  effects of scale invariance in galaxies:

\begin{enumerate}

\item In the past, the extent of galaxies was smaller than today. Galaxies were more compact
according to the ratio given by Equation (\ref{ratiogal}).

\item The cosmic expansion associated with scale invariance flattens the rotation curves as  galaxies age. 

\item In the past, the rotation curves should have steeper decrease and closer to the galactic center than today, thus suggesting  smaller fractions of dark matter in the past. This of course does  not preclude  other current effects of stellar dynamics to operate over the ages.   

\end{enumerate}

\subsection{Some Recent Results in Redshifted Galaxies} \label{redgal}
	
It is interesting to compare these theoretical results with some recent observations. Six star-forming galaxies with redshifts between  $z=0.8$ and 2.4 were analyzed  by Genzel et al. \cite{Genzel17}, and a large sample of 101 other galaxies from the KMOS and from the SINS/zC-SINF surveys with redshifts between $z=0.6$ and 2.6 were studied by Lang et al. \cite{Lang17}. Both surveys give trends  corresponding to the above point \#3: the observed past rotation curves were not flat, but were displaying steeply decreasing  velocity curves. Thus, even at distances of several  effective radii, the authors found that the dark matter fractions were modest  in the past. A variety of interpretations of the above results  have taken  place (see also  
\cite{Genzel20}). %MDPI: Please revise and provide reference citation (e.g.: [1]).
 However, the further analysis by Genzel et al. \cite{Genzel20} from the same survey confirmed that at $z \sim 2$ galaxies have very low dark matter fractions, below 20\%.
	
The observations  are also improving rapidly, and  very recently the above results were confirmed by new data from  different instruments and wavelengths.  In particular, a  hundred galaxies at the peak of their star formation ($z= 0.6 - 2.5$) were observed with ESO-VLT, LUCI, NOEMA at IRAM, and ALMA, and these were further analyzed, doubling  the previous samples \cite{Nestor Shachar23}. These new  observations show rather steep rotation curves, which---according to the authors---demonstrate that the fractions of dark matter strongly decrease with increasing redshifts: for example, at $z = 0.85$, the  fraction of dark matter is estimated to be 38\% of the typical value, and this fraction goes down to 17\% at $z= 2.44$.
	
\textls[-15]{In standard models with dark matter, the above decrease  of DM with redshifts is difficult to interpret. We do not know where, at high $z$,  the  dark matter is located, in order to enable such small dynamical signatures only. The possibility that the missing dark matter is lying in more external regions would be  a rather  troublesome one. Indeed, in the CDM theory  the baryons set in the potential wells already created by dark matter during the radiative era.  Thus,  it is not clear why galaxies would contain less dark matter in the past, since the accumulation of DM is considered preceding that of baryons. Another difficulty is that, contrarily to the above results showing less dark matter at higher $z$, some results on lensing, in particular for JWST-ER1,  seem to  suggest more dark matter for high-redshift galaxies, as will be seen in Section \ref{lensing}.
 The two above properties represent some  internal contradiction for the dark matter interpretation, and this tends to support our conclusion that the dark matter hypothesis is likely not the right solution for all problems in the dynamics of galaxies. }
	 
\subsection{The Connection between Dark Matter and Baryons: The Signature of a Gravity Effect}  \label{dmbar}
	
  There is  an interesting observation  about the radial acceleration  (RAR) studied by Lelli et al. \cite{Lelli17}, who showed that the relation between the  average accelerations $\frac{\upsilon^2}{R}$  and $\frac{GM}{R^2}$ for galaxies deviates from a 1:1 relation for gravities below $1.2 \times 10^{-8} $ cm s$^{-2}$. The RAR  is  accounted for by both MOND \cite{Lelli17} and by the  SIV theory \cite{MaedGueor20b}. As found by \cite{Lelli17} %MDPI: Please provide ref citation (e.g.: [1]).
 the amount of dark matter (DM, measured by the deviations from the 1:1 line) seems to be always directly determined by the amount of baryonic matter, although 2700 measurements have been made in different galaxies and {\emph{at different distances from their centers}}. Such a result is  pointing in favor of a gravity effect rather than for DM.
	
The same kind of relation between the  dynamical and  baryonic masses of central regions is also present in clusters of galaxies according to Chan (2022) \cite{Chan22}. Moreover, such a relation is also observed in redshifted galaxies, where Nestor Shachar et al. \cite{Nestor Shachar23} find a strong correlation between the fraction of DM and the baryon surface density within the effective radius (i.e., the part corresponding to the flat curve). The convergence of such  findings in three very different environments confirms the general  intimate  relation between  baryons and the assumed dark matter. It is clearly supporting  the views by Lelli et al.  that this general connection is the signature of gravity effect.

To summarize, the higher velocities in galaxy clusters, the flatter rotation curves at low redshifts than at high redshifts; the omnipresent connections between the distribution of baryons and of the supposed dark matter are all in support of a dynamical evolution at low densities governed by scale-invariant effects in the dynamics of galaxies.

%%%%%%%%%%%%%%%%%%%%%%%%%%%
\section{Wide Binary Stars}   
\label{wide}
%%%%%%%%%%%%%%%%%%%%%%%%%%%

The effects of scale invariance  should also be active in low-gravity systems with long timescales. An interesting case is that of very wide binary stars (wide binaries), where new interesting results from the astrometric Gaia satellite have recently come out. As is often the case for physical effects at the limit of experimental possibilities, this domain is rich in different and controversial views.
											
\subsection{Current Status of Observational Results on Very Wide  Binaries}  
\label{wideobs}

Among recent results, we  mention the study by Hernandez \cite{Hernandez22}, who studied a  sample  of  929 carefully selected binary pairs  from the Gaia Early Data Release 3 eDR3 \cite{GaiaColl21}  with high quality requirements. He found  that, below  $a_0 \simeq 1.2 \times 10^{-10}$ m s$^{-2}$, the velocities remain constant, independently of the orbital radius. This value $a_0$ corresponds to the  deep MOND limit, below which gravity behaves like   $g = \sqrt{g_{\mathrm{N}} \, a_0}$ (with $g_{\mathrm{N}}$  the Newtonian gravity), and orbital velocities remain constant according to \cite{Milgrom83,Milgrom09}. Another  study  was published by Hernandez \cite{Hernandez23} on the basis of the DR3 Gaia catalog with 800,000 binary candidates. Hernandez considered the binary probability $B_p$ for each star based on spectral, photometric and astrometric information, the RUWE index, the radial velocities, the isolation criteria,  the Gaia FLAME work package for mass estimate, etc. These very stringent quality  criteria gleaned a set of 450 top-quality isolated binaries, which do well to confirm the earlier   results%Please check that intended meaning has been retained
. 

Pittordis and Sutherland \cite{Pittordis23} studied a large sample of 73,159 wide binaries from the Gaia eDR3 objects within  300 pc and with a G magnitude brighter than 17.  With masses determined by a mass--luminosity relation, they analyzed the distributions of the relative projected velocities and found  a peak close to the Newtonian predictions, in addition to a long tail, likely produced by unseen objects. They noted, with some reserve, that standard gravity is preferred over MOND.

Chae \cite{Chae23} developed a Monte Carlo method to de-project the observed 2D data%Please check that intended meaning has been retained
. He proceeded to a calibration of the fraction $f_{\mathrm{multi}}$ of multiples  for gravities  $>10^{-10}$ m s$^{-2}$ and found a fraction of  0.3 to 0.5  in agreement with current results.The comparison of the kinematic acceleration $g= \upsilon^2/r$ and the Newtonian gravity shows a ratio  $g/	g_{\mathrm{N}}=1.43$ at $g_{\mathrm{N}}= 10^{-10.15}$ m s$^{-2}$ in agreement with one of the MOND generalizations. Some  critical remarks on other works are mentioned.

Banik, Pittordis, and Sutherland et al. \cite{Banik24} performed tests of the MOND dynamics with  wide binaries, from a highly selected sample of 8611 %MDPI: ` was removed, due to number having less than 4 digits.
 wide binaries from Gaia DR3 within 250 pc of the Sun. They  proceeded to a modeling of the orbits in the Newton and MOND approximations. Their statistical analysis   favored  a strong preference for Newtonian gravity, although their best model does not fully reproduce the observations. They also suggested further modifications of MOND.

A new study  by Chae \cite{Chae24} uses the catalog by El-Badry et al. \cite{El-Badry21} that provides a probability, $\Re$, of chance alignment. He applied, among others, the severe requirement that $\Re < 0.01$, i.e., mandating ``pure binaries''%Please check that intended meaning has been retained
.  He  pointed out that the estimate  $f_{\mathrm{multi}}$ by \cite{Banik24} of $f_{\mathrm{multi}}=0.70$ was incompatible with the range currently obtained. He  studied the  distributions of the sky-projected velocities as a function of the sky-projected separation $s$ for pure binaries.  Deviations from Newton's starts around $\log s=3.3$  (in au), increasing up to $\log s=3.8$, keeping about constant deviation by about 0.1 dex larger (Figure \ref{Chae13}), a property which appears to differ from MOND's. 

\subsection{Dynamical Evolution of Loose and Detached Systems}  
\label{lloo}

For increasing distances from a  star of  mass $M$, the attraction of the other star in a binary or of the galactic field  makes gravity vanish. This occurs at  the Jacobi radius, or Roche radius, where the galactic field becomes dominant; a value of $\sim$ 1.70 pc  is estimated for solar mass stars \cite{Jiang10}. Before this  limit, the dynamical acceleration  $\frac{\psi}{\tau_0} \times \upsilon $  may  overcome the   Newtonian gravity  $g_{\mathrm{N}}$. The equality of the two terms occurs at \cite{Maeder23}:
\begin{eqnarray} 
 g_0 \, = \, 
 % \frac{(1-\Omega_{\mathrm{m}})^2}{2}  \frac{4  \pi}{3} G \frac{3  H_0^2}{8 \pi G}  \frac{n c}{H_0} =
 \frac{(1-\Omega_{\mathrm{m}})^2}{4} \, n\,  c \, H_0\,  \quad \quad 
 \mathrm{or} \quad  g_0 \,= \, \frac{n\, c \, (1-\Omega_{\mathrm{m}}) (1-\Omega^{1/3}_{\mathrm{m}})}{2\, \tau_0}\, .
 \label {go}
 \end{eqnarray}

The product $c \times H_0$ is equal to $6.80 \times 10^{-8}$ cm s$^{-2}$. For $\Omega_{\mathrm{m}}$=0, 0.10, 0.20, and  0.30, we obtain $g_0 =$ (1.70, 1.36, 1.09, 0.83) $ \times \, 10^{-8}$ cm s$^{-2}$, respectively. These values of $g_0$ cover the domain  of $a_0$. For gravities below $g_0$, the equation of motion (\ref{Nvec4}) accepts, {\emph{as an approximation}},   a relation of the form $g=\sqrt{g_{\mathrm{N}} g_0} \;$ \cite{Milgrom83}, as discussed above; it is 
 valid   over the last  400 Myr to an accuracy of 1-2\% depending on $\Omega_{\mathrm{m}}$ \cite{Maeder23}. For a given total mass $M_{\mathrm{tot}}$ of a binary, the equality of the two terms occurs at a separation:
\begin{equation}
s_{\mathrm{trans}}\, = \, \sqrt{\frac{G \,M_{\mathrm{tot}}}{g_0}} \,, \quad \mathrm{for \;M_{\mathrm{tot}}= 1.0\;  M_{\odot}, 
\;  s_{\mathrm{trans}}= 7.0\; [kau] \; or\; 0.034\; [pc]. }
\label{strans}
\end{equation}

What happens for wide binaries  near the transition? For a decrease in effective gravity,  the eccentricity $e$ and  radius  $r_0$ \, of the circular orbit  are tending to $\infty$ from Equations (\ref{traj}) and (\ref{r0})%Please check that intended meaning has been retained
. The orbit  extends  with an increase in the ellipticity up to $e=1$, where it becomes parabolic.  At this limit, we have
\begin{equation}
C \, = \,  \frac{GM\, \psi^2}{\mathcal{L}^2 \, \tau^2_0}      \, \equiv   \,\frac{1}{r_0}.
\label{Climit}
\end{equation}

For a further decline in gravity, $e$  becomes larger than 1 and the orbit is hyperbolic, tending to a straight line. Let us examine the behavior of the orbital velocity at this limit:
%

%\begin{adjustwidth}{-\extralength}{0cm}
%\centering %% If there is a figure in wide page, please release command \centering
\begin{equation}
 \frac {d^2 \bf{r}}{d \tau^2}  \, =  \frac{\psi(\tau)}{\tau_0}   \frac{d\bf{r}}{d\tau} \,,    \quad \quad \mathrm{or} \; \; 
 \frac{d\upsilon}{\upsilon}  \,= \,   \frac{\psi(\tau)}{\tau_0} \; d\tau\, \,= \, -  \frac{\dot{\psi}}{\psi} \, d\tau\, = \, -  \frac{d\psi}{\psi},  \quad \quad
\mathrm{with} \; \dot{\psi} =-\psi^2/\tau_0.
\label{Nvec5}
\end{equation}
%\end{adjustwidth}
%

%\begin{adjustwidth}{-\extralength}{0cm}
%\centering %% If there is a figure in wide page, please release command \centering
\begin{equation}
\mathrm{Integration\; gives} \quad \ln \upsilon = - \ln \psi + const.  \quad \quad \mathrm{thus }   \; \; \upsilon =  \frac{k}{\psi}= 
\frac{k}{(1-t_{\mathrm{in}})}\, [t_{\mathrm{in}}+\frac{\tau}{\tau_0}(1-t_{\mathrm{in}})]\,,
\label{vt}
\end{equation}
%\end{adjustwidth}
%
\textls[-15]{where $k$ is a constant. Setting $\tau=\tau_0$, we have $\upsilon(\tau_0)=\frac{k}{(1-t_{\mathrm{in}})} $ and $\upsilon(\tau)$ becomes\linebreak  $\upsilon(\tau) = \upsilon(\tau_0)\,  [t_{\mathrm{in}}+\frac{\tau}{\tau_0}(1-t_{\mathrm{in}})].$ The velocity no longer remains constant with secular variations as in the bound system, but it  increases linearly in time $\tau$ under the effect of the dynamical acceleration $ \frac{\psi(\tau)}{\tau_0} \; \upsilon$.  Turning to the distance, $r$, covered during this motion, we have from (\ref{vt}),}
\begin{equation}
\frac{dr}{d\tau} = \upsilon(\tau_0)\,  [t_{\mathrm{in}}+\frac{\tau}{\tau_0}(1-t_{\mathrm{in}})], \quad \mathrm{thus} \; \;
r(\tau) = \frac{1}{2} \frac{\tau_0}{(1-t_{\mathrm{in}})}\, \upsilon(\tau_0)\, [t_{\mathrm{in}}+\frac{\tau}{\tau_0}(1-t_{\mathrm{in}})]^2.
\label{rt}
\end{equation}

Thus, when the dynamical gravity $\frac{\psi}{\tau_0}  \upsilon$  dominates, the motion becomes hyperbolic, with $e>1$, a velocity increasing  linearly with time and  a  covered distance   with a quadratic dependence in time, $r(\tau) =r(\tau_0) [t_{\mathrm{in}}+\frac{\tau}{\tau_0}(1-t_{\mathrm{in}})]^2.$ We call this regime the \emph{dynamical regime}. In order to numerically estimate the  changes in velocities and covered distances, we consider a reference time $\tau_{\mathrm{N}}$   marking the transition of the Newtonian  regime and to the dynamical one. We then obtain the following ratios:
\begin{equation}
\frac{\upsilon(\tau_0)}{\upsilon(\tau_{\mathrm{N}})}= \frac{1}{ [t_{\mathrm{in}}+\frac{\tau_{\mathrm{N}}}{\tau_0}(1-t_{\mathrm{in}})]},
\quad \quad \mathrm{and}\; \; 
\frac{r(\tau_0)}{r(\tau_{\mathrm{N}})}= \frac{1}{ [t_{\mathrm{in}}+\frac{\tau_{\mathrm{N}}}{\tau_0}(1-t_{\mathrm{in}})]^2}.
\label{vn}
\end{equation}

This gives the change in velocity and radius, between the transition time  $\tau_{\mathrm{N}}$ and  today in the dynamical regime, for binary systems in the Milky Way.

\subsection{Comparison with the Analysis by Chae \cite{Chae24}} 
\label{fchae}

\vspace{-12pt} 
\begin{figure}[H]

\includegraphics[width=0.7\textwidth]{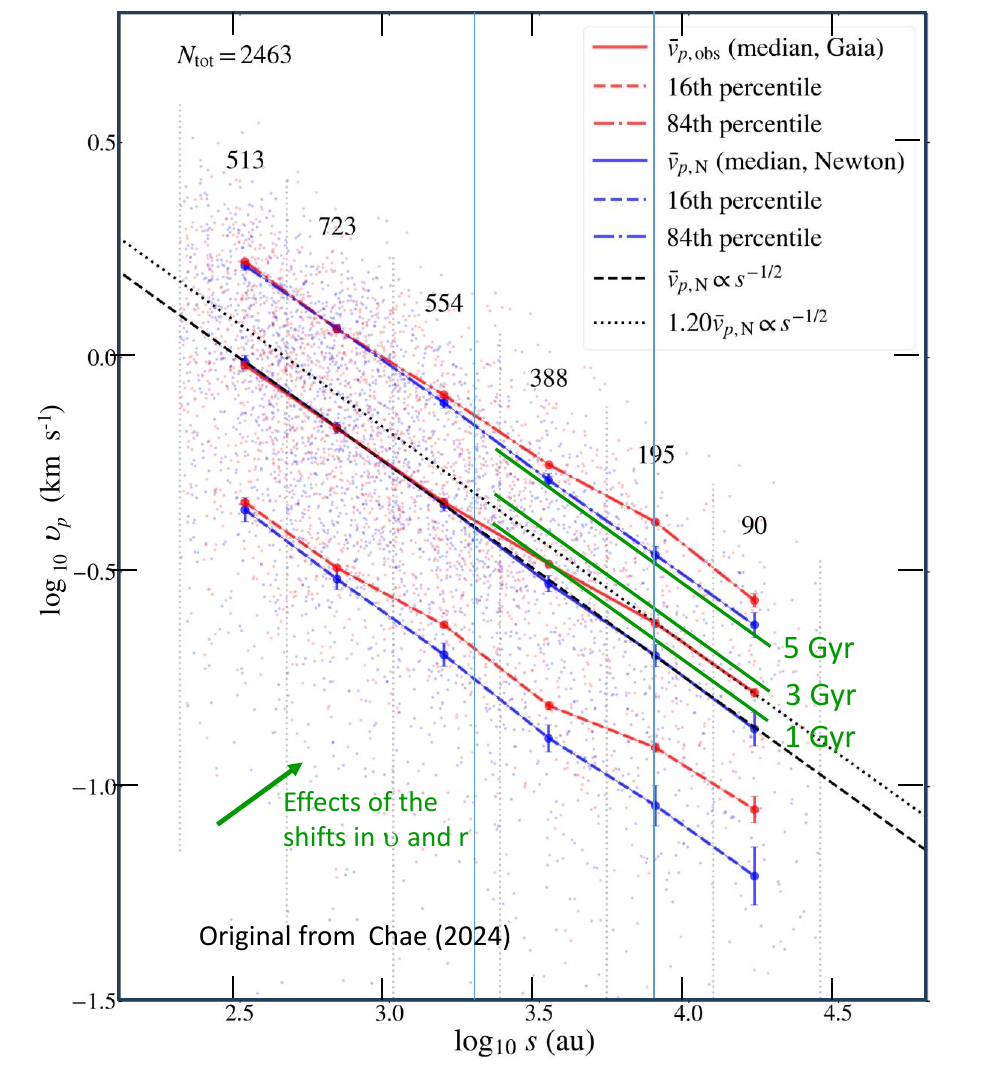}
\caption{Projected velocities $\upsilon_p$ as a function of separation $s$ for the main sample in Figure 13 %MDPI: Please confirm that figure citation does not refer to any figure from within the manuscript and it should not be linked. Figure was moved after its first citation. Please check and confirm.
 from Chae \cite{Chae24}. The very small red dots are the observed values and the blue dots are the Newtonian values in the Monte-Carlo simulations. The larger dots, red (obs.) and blue (simul.) and their connecting lines are the medians and percentiles, as indicated. The central black dashed line shows the Keplerian relation in $1/\sqrt{s} $. The green lines are additional indications of the mean deviations along the hyperbolic path of the loose systems after 1, 3, and 5 Gyr from the time, $\tau_N$, the transition from Newtonian to dynamical acceleration. The direction of the effects in velocity and separation is indicated by a green arrow at the bottom left. The deviation of a given system increases linearly with time as indicated by Equation (\ref{vn}). The mean observed relation for the loose systems corresponds to an   evolution during about  2 to 3 Gyr. In  a few Gyrs, loose systems drift away from the Newtonian relation in a way compatible with the dynamical evolution.  (Figure adapted from Chae \cite{Chae24}).}
\label{Chae13}
\end{figure}

\vspace{22pt}

Figure \ref{Chae13} shows in red  the  relation between $\log \upsilon$ and $\log s$ for a sample of 2463 binary pairs with masses derived from the absolute G-band magnitude, $M_G$, see comments. From an observed  set of  $M$ and $s$ values, each Monte-Carlo (MC) realization draws the velocity from the classical Newtonian equations with the appropriate statistical distributions of the orbital parameters. Series of 400 MC simulations are performed to estimate statistical error bars (blue line). The average total mass of the binary systems is around 1.5 M$_{\odot}$; thus,  the main sequence lifetimes are of several Gyr. We consider systems starting  deviating from Newton's Law at 1, 3 ,and 5 Gyr ago, i.e., at past ages $\tau_{\mathrm{N}}=$ 12.8,  10.8, and  8.8  Gyr. Thus, from Equation (\ref{vn}), the shifts  in velocities and distances since time $\tau_{\mathrm{N}}$, respectively, are:
\begin{equation}
\log \left[\frac{\upsilon(\tau_0)}{\upsilon(\tau_{\mathrm{N}})}\right]=0.020, 0.064, 0.113, \quad \quad  \mathrm{and} \quad
\log \left[\frac{r(\tau_0)}{r(\tau_{\mathrm{N}})}\right]=0.040, 0.128, 0.226.
\label{val}
\end{equation}

\hspace{22pt}In Figure \ref{Chae13}, these shifts produce  displacements from the red to the green lines, for binaries at different locations on the reference line $\upsilon \sim 1/\sqrt{s}$. The mean shift increases between $s= 3$ [kau]  and 8  [kau],  in agreement with $s_{\mathrm{trans}}$. The observed shifts correspond to  an evolution in the free regime for  about 2 to 3 Gyr. These shifts are clearly larger than the error bars. We note that all the possible variations/combinations of the selection criteria in the sampling by \cite{Chae24} give rather similar shifts.

\begin{figure}[H]

\includegraphics[width=0.8\textwidth]{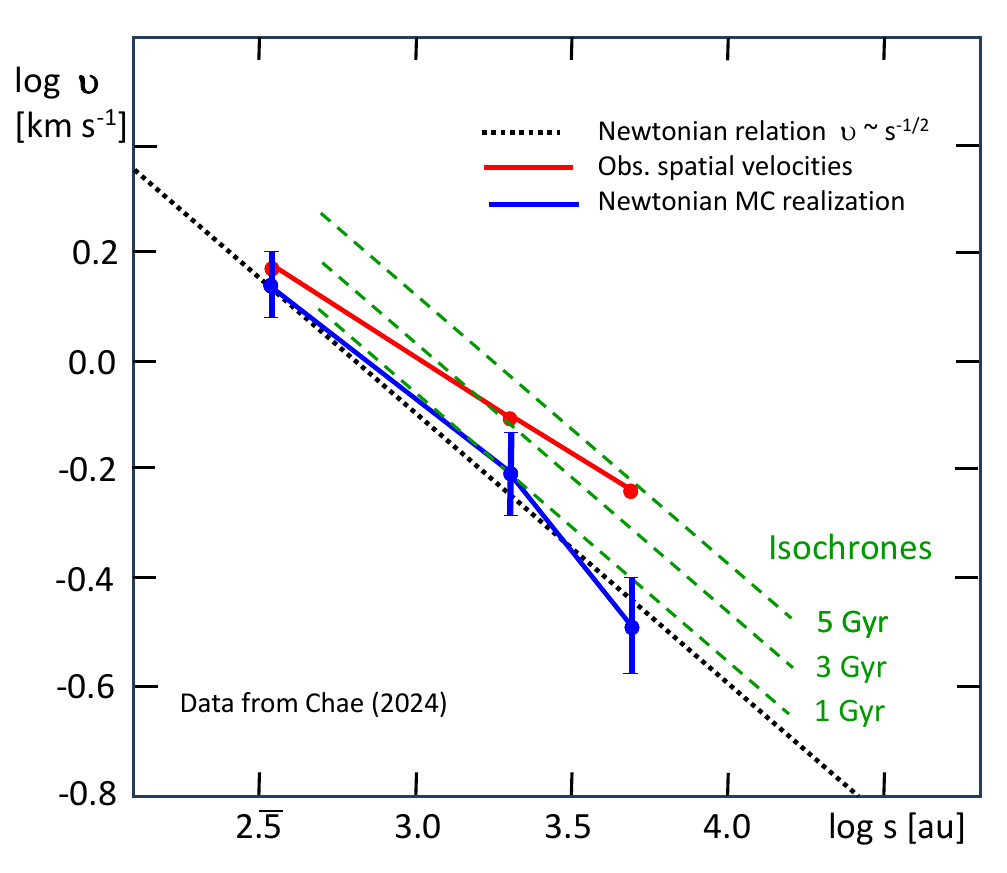}
\caption{Comparisons %MDPI: Figure was moved after its first citation. Please check and confirm.\hl{} %MDPI: Please change the hyphen (-) into minus sign ($-$, "U+2212"). e.g., "-1" should be "$-$1".
 of observations and theory for a sample of 40 very wide binaries with exceptionally precise radial velocities, with individual relative errors smaller than 0.005, as selected by Chae \cite{Chae24}.  The green broken lines shows the isochrones corresponding to departures from the Newtonian law after 1, 3, and 5 Gyr of evolution under the dynamical acceleration in the scale-invariant theory. The departure from Newton's relation is progressive and tends towards a value between 3 and 5 Gyr (data are from Chae \cite{Chae24}).}
\label{Chae21}
\end{figure}

In Figure \ref{Chae21}, Chae \cite{Chae24} analyzed a smaller sample including only the 40 binaries with uncertainties $< 0.5$ \%, comparable to  astrometry. Three bins are formed; the highest $s$-bin shows a large deviation from Newtonian relation, compatible with  free motion during about 5 Gyr. This extreme sample supports the existence of significant deviations from Newtonian relation for $\log s \geq 3.5$ and above. It appears as a robust conclusion that  the direction and size of the observed effects are compatible with a dynamical  evolution  dominated by the additional scale invariance term  for several Gyr. 

\subsection{Comparison with the Analysis by Hernandez \cite{Hernandez23}}

\begin{figure}[H]

\includegraphics[width=0.75\textwidth]{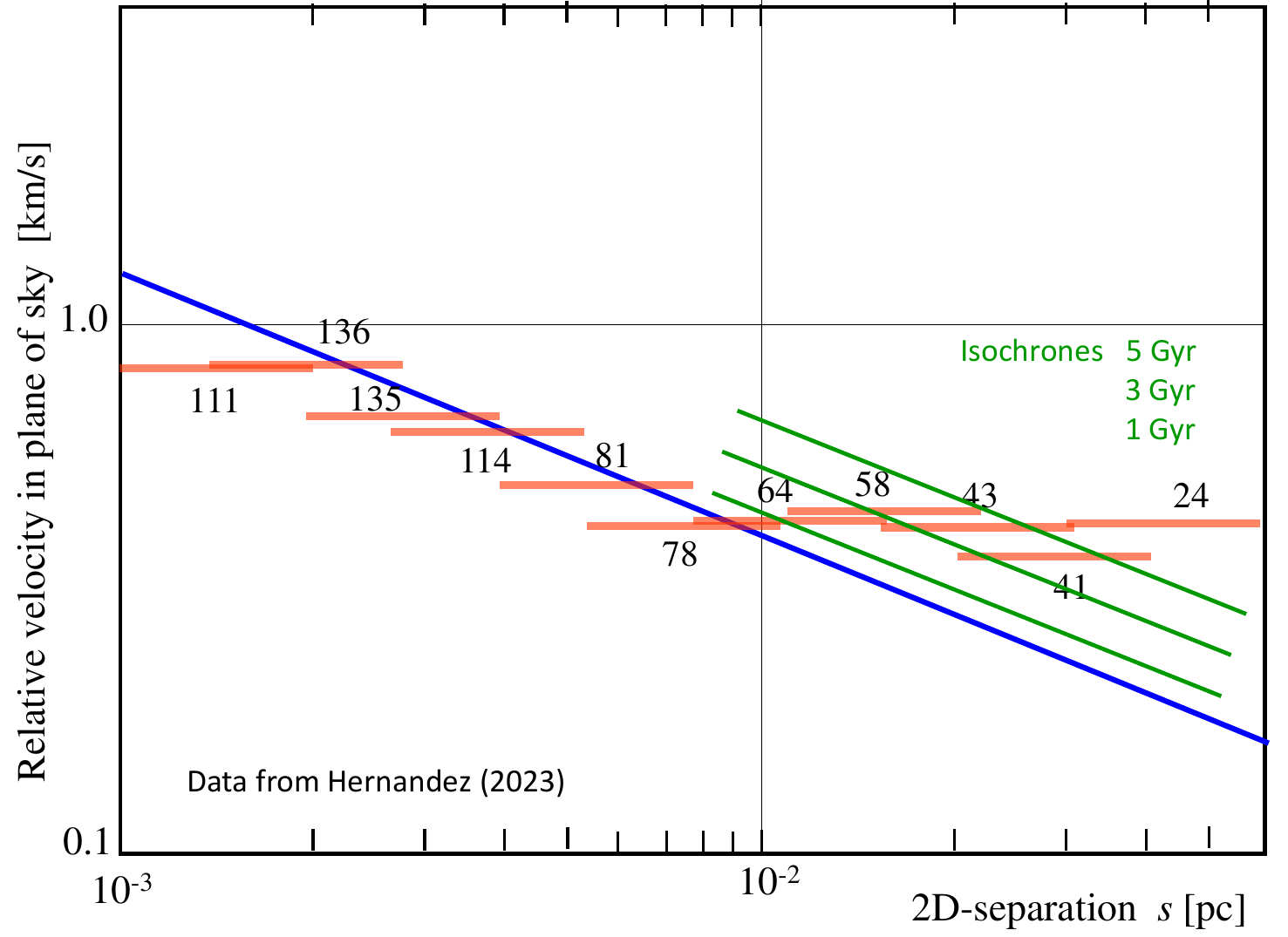}
\caption{Binned %MDPI: Please change the hyphen (-) into minus sign ($-$, "U+2212"). e.g., "-1" should be "$-$1".
 root mean square $\langle \Delta V \rangle$ of the relative velocities in the plane of sky as a function of the 2D projected separation $s$ for 450 binary pairs by Hernandez \cite{Hernandez23}. There is a partial overlap of the  binned  pairs. The number of binary pairs in the various means are indicated. The green  lines show the  isochrones corresponding to departures from the Newtonian law after 1, 3, and 5 Gyr of evolution (adapted from \cite{Hernandez23}).}
\label{Hernan23}
\end{figure}

\hspace{22pt}The sample of 450 wide binaries  by Hernandez \cite{Hernandez23} results from  stringent quality criteria (Section \ref{wideobs}). Mass estimates make use of  an internal Gaia FLAME work-package based spectroscopic information in DR3. The relative velocity $\Delta V$ analysis  in  the plane of the sky was  primarily determined  through Gaia \emph{proper motions in two perpendicular directions}. The binned  root-mean-square  $\langle \Delta V \rangle$ of the relative velocities between the two components of selected pairs are shown in Figure \ref{Hernan23} as a function of the 2D projected separations $s$  between the two components of each binary pair.

\hspace{22pt}Figure \ref{Hernan23} shows the results. The  blue line shows the Newtonian relation $\upsilon \sim s^{-1/2}$ from  simulations of 50,000 binary pairs  of two  1 M$_{\odot}$ mass \citep{Jiang10}. This line  follows a Keplerian relation $\langle \Delta V \rangle \sim  s^{-1/2}$ up to the Jacobi radius, well beyond the value of about 0.01 pc (note that 0.01 pc corresponds to 2.063 kau).

\hspace{22pt}The Keplerian relation is very well followed for separations up to about  $10^{-2}$ pc. However, for larger separations, the mean relative velocities significantly deviate from the Keplerian relation and tend toward an essentially flat constant value (see Figure \ref{Hernan23}). 

\hspace{22pt}The comparison of the results by  Hernandez \cite{Hernandez23} with  isochrones showing the duration of the free evolution  also  supports an evolution in the \emph{dynamical regime}  of the order 5 Gyr,  in agreement with the best-quality results in Figure \ref{Chae21}. The key point is that these different results on very wide binaries are  supporting significant deviations from the standard Newtonian predictions, consistent with an evolution dominated by Equation (\ref{rt}) for several Gyr.

\hspace{22pt}\textls[-25]{A possible difficulty has been raised by Loeb \cite{Loeb22}: in  wide binaries, the gravitational redshift has  about the same amplitude as  the Doppler--Fizeau effect. This might be a problem; however, it is not necessarily a problem if we  consider the differential effects. Following Loeb, we express the velocity shift by   gravitation $\upsilon_{\mathrm{GR}} = - \frac{G\, M}{cR}$; with the mass--radius relation $\frac{R}{R_{\odot}} \approx  \left(\frac{M}{M_{\odot}} \right)^{0.8}$ for solar-mass  stars, it becomes $\upsilon_{\mathrm{GR}}  \approx - 0.6  \left(\frac{M}{M_{\odot}}\right)^{0.2} $%Please check that intended meaning has been retained
	. For a mean  a mean mass ratio $M_2/M_1$ of about 0.5 to power of 0.2, this  gives  a value of 0.87 for  the ratio of the gravitational redshifts of the two components, i.e., typically a velocity  difference of about 0.08 km/s which is small for a mean velocity of 0.60 km/s. Thus, the differences of gravitational redshifts between the two components are much smaller than the velocity shifts. Furthermore, in the data by Hernandez, the relative motions  are mainly derived from  the proper motions in the plane of the sky,  where gravitational effects are not in action%Please check that intended meaning has been retained
	. Moreover, gravitational redshifts do not introduce any  systematic bias in the mean of many binaries.}

\hspace{22pt}As a conclusion, the recent results from Gaia DR3 for very wide binaries by \cite{Hernandez23} and \cite{Chae24} show orbital velocities larger than Newtonian  for separations larger than  about  3 kau, but they are in agreement  with the predictions of the scale-invariant theory. It seems difficult to advocate for a halo of dark matter surrounding each binary system, as currently performed in the domain of galaxies and clusters.\\

%%%%%%%%%%%%%%%%%%%%%%%%%%%%%%%
\section{The Discrepancy between the Lensing and the Spectroscopic Mass of Galaxies} 
\label{lensing}
%%%%%%%%%%%%%%%%%%%%%%%%%%%%%%%

\hspace{22pt}Gravitational lensing is a powerful tool of astrophysics. It is no exception to the current problems related to dark matter, in the sense that the masses from lensing  are generally larger those from photometry,
 see Sections \ref{JW} and \ref{SLACS}.  Thus, it is worth examining the effects of scale invariance in the context of lensing.

\subsection{The Geodesic of Deflected Light Rays in the Scale-Invariant Theory}  \label{geodlight}

\hspace{22pt}To calculate the modified trajectories  of a body  near another  mass,  the starting point is  the equation of the geodesic \cite{Dirac73,BouvierM78}:
\begin{equation}
\frac{du^{\alpha}}{ds}+ \Gamma^{\alpha}_{\mu \nu} u^{\mu} u^{\nu} -\kappa_{\mu}u^{\mu} u^{\alpha}+ \kappa^{\alpha} = 0.
\label{geod}
\end{equation}

\hspace{22pt}For the metric  of the system, we adopt the Kottler metric \cite{Kottler18}:
\begin{equation}
ds^2= \frac{-dr^2}{1-\frac{2 GM}{rc^2}-\frac{\Lambda r^2}{3}} -r^2(d\vartheta^2+\sin^2 \vartheta d\varphi^2)+
\left(1-\frac{2 GM}{r c^2}-\frac{\Lambda r^2}{3}\right) c^2 dt^2.
\label{Sch}
\end{equation}

\hspace{22pt}$\Lambda = \lambda^2 \Lambda_{\mathrm{E}}$, with$\Lambda_{\mathrm{E}}$ the cosmological constant in GR; $ \lambda$  is the scale factor,  a slowly variable universal function. We  calculate  the geodesic  for a body of mass $m=GM/c^2$; the case of light rays is specified later. The  geodesic differs from the classical case of GR by the terms  $ -\kappa_{\mu}u^{\mu} u^{\alpha}+ \kappa^{\alpha}$:
\begin{eqnarray}
\alpha =1 &:& \quad -\kappa_{\mu}u^{\mu} \frac{dr}{ds} \quad \rightarrow \quad -\kappa_0 \frac{dt}{ds} \frac{dr}{ds}, \\ \nonumber
\alpha =2 &:& \quad -\kappa_{\mu}u^{\mu} \frac{d\theta}{ds} \quad \rightarrow \quad  -\kappa_0 \frac{dt}{ds} \frac{d\theta}{ds}, \\ \nonumber
 \; \alpha =3 &:& \quad -\kappa_{\mu}u^{\mu} \frac{d\phi}{ds} \quad \rightarrow \quad -\kappa_0 \frac{dt}{ds} \frac{d\phi}{ds}, \\ \nonumber
 \alpha =0 &:&  \quad -\kappa_{\mu}u^{\mu} \frac{dt}{ds} + \kappa_0 \quad \rightarrow \quad  -\kappa_0 \frac{dt}{ds} \frac{dt}{ds}+\frac{\kappa_0}{c^2}. 
\end{eqnarray} 

\hspace{22pt}With $\kappa_0= -\dot{\lambda}/\lambda= 1/t$. We adopt a simplified form of the metric:
\begin{eqnarray}
ds^2  =  -e^{\lambda} dr^2-r^2 (d\theta^2+\sin^2 \theta d\phi^2)+e^{\nu} c^2  dt^2;  \; \nonumber \\ \mathrm{with}\quad
e^{\lambda}=  \frac{1}{1-\frac{2 m}{r}-\frac{\Lambda r^2}{3}}, \quad
 e^{\nu}=\left(1-\frac{2 m}{r}-\frac{\Lambda r^2}{3}\right) ,
\label{sc}
\end{eqnarray} 
where  $m= GM/(c^2)$  has the units of a length and  is subject to slow time-variations  according to Equation (\ref{drdt}). We also assume, as usual, that the motions occur in a plane with  $\theta=\pi/2$, and we obtain

\begin{eqnarray}
\frac{d^2 r}{ds^2}+\frac{1}{2} \frac{d\lambda}{dr}\left(\frac{dr}{ds}\right)^2- r e^{-\lambda}\left(\frac{d\phi}{ds}\right)^2 +
\frac{e^{\nu-\lambda}}{2}\frac{d\nu}{dr}\left(\frac{c^2 \, dt}{ds}\right)^2  
-\kappa_0 \frac{dt}{ds} \frac{dr}{ds} = 0,  \label{d2r} \nonumber \\
\frac{d^2\phi}{ds^2}+\frac{2}{r} \frac{dr}{ds} \frac{d\phi}{ds}  -\kappa_0 \frac{dt}{ds} \frac{d\phi}{ds}  =  0,  \quad  \quad \mathrm{and} \;   \;
\frac{d^2 t}{ds^2}+ \frac{d \nu}{ds}\frac{dt}{ds}-\kappa_0 \left(\frac{dt}{ds}\right)^2+\frac{\kappa_0}{c^2}  =  0, \label{et}
\end{eqnarray}

\hspace{22pt}The last two equations may be easily integrated. Let us call $\upsilon= \frac{d\phi}{ds}$; for the first one, we obtain
\begin{equation}
\frac{d\upsilon}{ds}+ \frac{2}{r} \frac{dr}{ds} \upsilon -\frac{1}{t}\frac{dt}{ds} \upsilon=0, \quad \quad \mathrm{and} \; \;
\frac{d\upsilon}{\upsilon}+ 2 \frac{dr}{r}- \frac{dt}{t}=0,\; \;\quad  \mathrm{giving}\; r^2 \, \frac{d\phi}{ds} = h t.
\label{ht}
\end{equation}
Here, $h$ is an integration constant,   representing the angular momentum (divided by $c$) per mass unit. This  is the modified conservation law, which increases with time $t/t_0$. This is  consistent with  the scale-invariant two-body problem \cite{MBouvier79,Maeder17c}, where the orbital radius slowly increases, while the orbital velocity $r \, \frac{d\phi}{ds}$  keeps constant,  see Section \ref{Sinet} . Turning to the third equation of (\ref{et}),  we  multiply it by $\left(\frac{ds}{dt}\right)^2$ and obtain
\begin{equation}
\frac{d^2t}{dt^2}+\frac{d\nu}{dt} \, \frac{dt}{dt}- \kappa_0+\frac{\kappa_0}{c^2} \, \left(\frac{ds}{dt}\right)^2= 0.
\end{equation}
The first two terms are zero, the second one because $\nu$ is scale-invariant, independent of time, since in the  parenthesis of (\ref{Sch}), both mass and radius  are scaling like $1/\lambda$,  so that $2m/r$ is scale-invariant; the same is the case for  $\Lambda$,  scaling like $\lambda^2$. The two terms with $\kappa_0$ are also equal to zero  in the $t$-equation. Thus,   one is just brought back to the classical case; calling  upon $w=ds/dt$,  one is left with%Please check that intended meaning has been retained

\begin{equation}
\frac{dw}{ds}+  \frac{d\nu}{ds} w =0,  \quad \quad \mathrm{with\; solution} \; \; w=dt/ds= k \, e^{-\nu(s)}, \;\; k=const.
\end{equation}

\hspace{22pt}Now,  we turn to the more difficult Equation (\ref{d2r}). Let us quote \cite{Eddington23}, who said about the corresponding  equation in the GR context, {\emph{``Instead of troubling to integrate (\ref{d2r}) we can use in place of it  (\ref{sc})}}'' (the labels of both equations have been changed appropriately). Thus, from (\ref{sc}) we obtain
\begin{equation}
e^{\lambda} \left(\frac{dr}{ds}\right)^2 + r^2  \left(\frac{d\phi}{ds}\right)^2 - 
e^{\nu} c^2 \left(\frac{dt}{ds}\right)^2 +1=0.
\label{s3}
\end{equation}
With the above expression for $w$, the third term can also be written 
$
-e^{\nu} c^2 k^2 e^{-2\nu}=  -c^2 k^2 e^{-\nu}=
 -\frac{c^2 k^2}{[ 1-\frac{2 GM}{r c^2}-\frac{\Lambda r^2}{3}]}.
$
Thus, Equation (\ref{s3}) becomes
\begin{eqnarray}
\left(\frac{dr}{ds}\right)^2 +  r^2  \left(\frac{d\phi}{ds}\right)^2  -\frac{2m}{r} \left(1+ r^2  (\frac{d\phi}{ds})^2 \right) 
-\frac{\Lambda r^2}{3}\left(1+ r^2  (\frac{d\phi}{ds})^2\right)= c^2 k^2-1.
\label{B1}
\end{eqnarray}
Eliminating $(\frac{d\phi}{ds})$ with the last expression of (\ref{ht}), we obtain
\begin{equation}
\left(\frac{ht}{r^2} \frac{dr}{d\phi}\right)^2 +\frac{h^2 t^2}{r^2}=c^2k^2 -1+ \frac{2m}{r}+ 
\frac{2mh^2 t^2}{r^3}+\frac{\Lambda r^2}{3} +\frac{\Lambda    h^2 t^2}{3}.
\end{equation}
We divide it by $h^2 t^2$ and set $u=1/r$ for consistency with the classical Binet equation  given by Equation (\ref{Binet}):

\begin{equation}
\left(\frac{du}{d\phi} \right)^2 +u^2 =\frac{c^2 k^2 -1}{h^2 t^2}+ \frac{2mu}{h^2 t^2}+ 
2 mu^3 +\frac{\Lambda}{3 u^2 h^2 t^2} +\frac{\Lambda}{3}.
\end{equation}

\hspace{22pt}Now,  we derive this equation with respect to $\phi$ and obtain
\begin{eqnarray}
2 \left(\frac{du}{d\phi}\right) \frac{d^2u}{d\phi^2}  +2 u \frac{du}{d\phi} =
\frac{c^2 k^2 -1}{ h^2} \frac{d}{d\phi} \left(\frac{1}{t^2}\right)  + \frac{2m}{h^2 t^2}  \frac{du}{d\phi}+ 
+\frac{2u}{h^2 t^2} \frac{dm}{d\phi}+ 
  \frac{2mu}{h^2} \frac{d}{d\phi}\left(\frac{1}{t^2} \right) +\nonumber\\
6 mu^2 \frac{du}{d\phi}+2 u^3 \frac{dm}{d \phi}  
-2 \frac{\Lambda_{\mathrm{E}} \lambda^2}{3  h^2 t^2} \frac{1}{u^3}\frac{du}{d\phi}+
\frac{\Lambda_{\mathrm{E}} }{3  h^2 u^2} \frac{d}{d\phi} \left(\frac{\lambda^2}{t^2}\right)
+\frac{2\Lambda_{\mathrm{E}} \lambda}{3} \frac{d\lambda}{d\phi}.
\label{compl}
\end{eqnarray}

\hspace{22pt}There is  a direct relation between $\phi$ and time since the time $t$ is uniquely defined  along the geodesic described by the variation of angle $\phi$. Thus, $\lambda$,  which is a function of time $t$, is also a  function of $\phi$. Let us now multiply the result by $\frac{1}{2} \frac{d \phi}{du}$:
\begin{eqnarray}
\frac{d^2u}{d\phi^2}  +u & = &\frac{c^2 k^2 -1}{2 h^2} \frac{d}{du} \left(\frac{1}{t^2}\right)  + \frac{m}{h^2 t^2}+ \frac{u}{h^2 t^2} \frac{dm}{du}+
\frac{mu}{h^2} \frac{d}{du}\left(\frac{1}{t^2} \right) \nonumber \\
& + &3 mu^2+u^3 \frac{dm}{du} -\frac{\Lambda_{\mathrm{E}} \lambda^2}{3  h^2 t^2} \frac{1}{u^3}+
\frac{\Lambda_{\mathrm{E}} }{6  h^2 u^2} \frac{d}{du} \left(\frac{\lambda^2}{t^2}\right)
+\frac{\Lambda_{\mathrm{E}} \lambda}{3} \frac{d\lambda}{du},
\label{compl}
\end{eqnarray}

\hspace{22pt}This equation combines  the effects of general relativity and scale invariance. It can be used to study the advance of perihelion and other post-Newtonian effects in the scale-invariant context.

\subsection{The Invariance of Lensing to Scale Transformations}  

\hspace{22pt}For a light ray, we have the  additional condition  $ds=0$, {{i.e.,}} its proper time never changes. According to Equation (\ref{ht}), this implies $h=\infty$, see \cite{Eddington23}.  This simplifies Equation (\ref{compl}), where only three terms are remaining on the right side. From the gauging condition   $3 \,\frac{ \dot{\lambda}^2}{\lambda^2} \, = \,c^2 \lambda^2 \,\Lambda_{\mathrm{E}} $ \cite{Maeder17a,MaedGueor23} (with $c^2$ for the consistency of the units),  we obtain, with $\lambda=t_0/t$ in the units of the cosmological models, $\Lambda_{\mathrm{E}} = \frac{3}{c^2 \, t^2_0}$%Please check that intended meaning has been retained
;  in current units, $\Lambda_{\mathrm{E}}=  \frac{3}{c^2 \, \tau^2_0}$ with $\tau_0$ the age of the Universe (see also \cite{MaedGueor21a}  for a discussion of these
properties).  Thus, Equation (\ref{compl}) becomes
\begin{equation}
\frac{d^2u}{d\phi^2}  +u =  3 mu^2+u^3 \frac{dm}{du}+ \left(\frac{1}{c^2   \tau^2_0}\right) \lambda \frac{d\lambda}{du}.
\label{BM}
\end{equation}

\hspace{22pt}The scale factor $\lambda$ is subject to the current  limitations   (Section \ref{th}). If $\lambda$ is a constant, we again have the  usual  equation for lensing in GR. $c \tau_0$ is a distance of the order of the radius of the Hubble sphere $c/H_0$. Numerical integration of the scale-invariant geodesics, for the specific case of the $z_L \cong 2.0$ lens galaxy in the extreme system JWST-ER1 (Section \ref{JW}) shows that the two additional terms  are  negligible, with about $\mathcal{O}$($10^{-6}$) of the Einstein term. Thus, we may conclude that the photon's geodesics remain unchanged and the deflection angle $\alpha= 4 G M/(c^2R)$  is not affected by scale-invariant effects, noting also that the mass $M$
and radius $R$ vary the same way with time.

\subsection{The Problem of the Photometric Masses: The Case of JWST-ER1}  \label{JW}
\begin{figure}[H]%[t!]

\includegraphics[width=12cm, height=8cm]{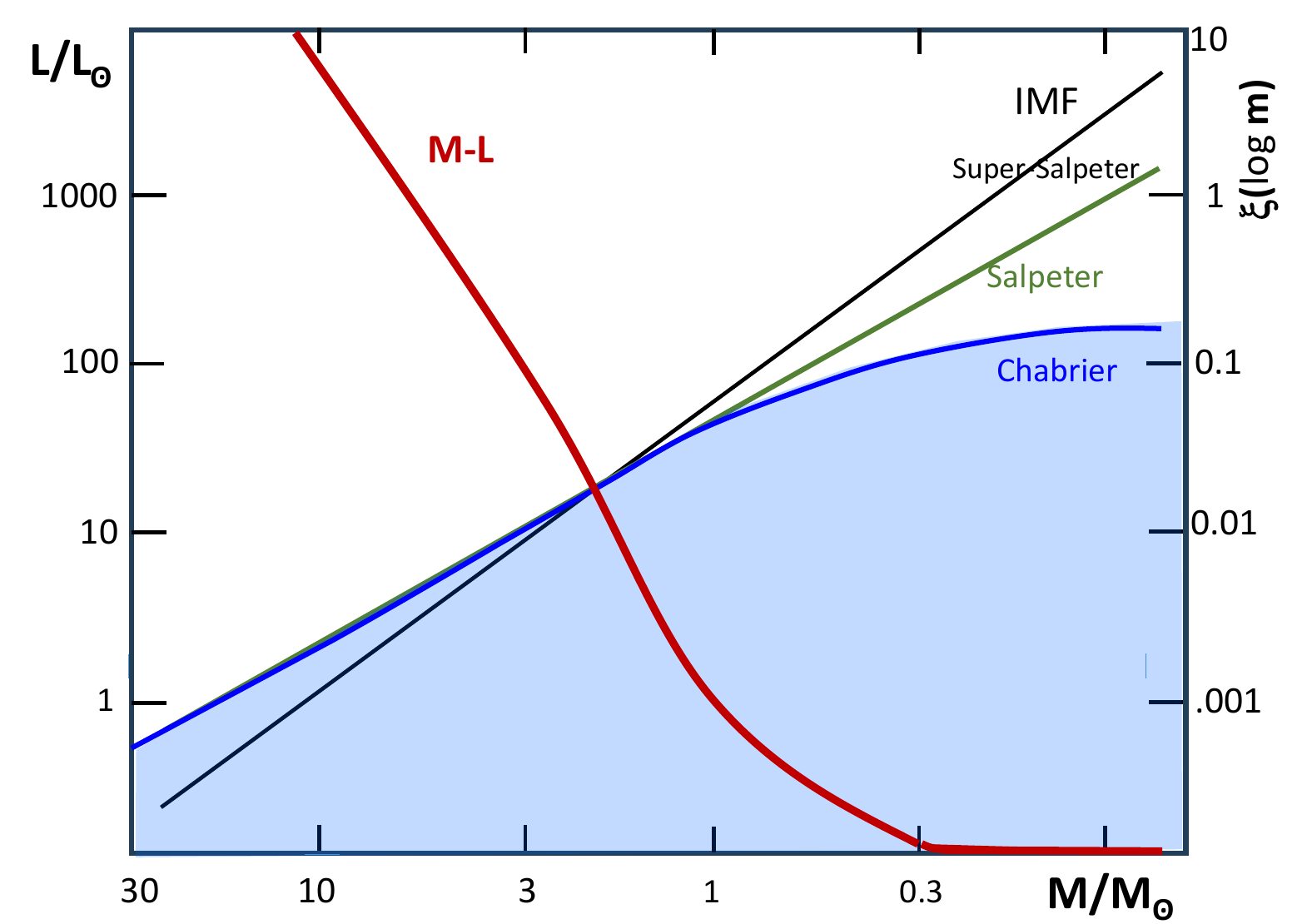}
\caption{The %MDPI: Figure was moved after its first citation. Please check and confirm.
 red thick curve represents the mass--luminosity relation on the zero-age sequence \cite{Maeder09}, according to the scale indicated on the left vertical axis. The other curves describe various IMF:  Chabrier \cite{Chabrier03},  Salpeter with $x=1.35$ \cite{Salpeter55}, and    the so-called Super-Salpeter by \cite{Dokkum23},  according to the scale indicated on the right vertical axis. Note that, in $\xi(\log m)$, the log is a decimal. The Figure is inspired by \cite{Dokkum23}.
}
\label{imf}
\end{figure}

\hspace{22pt}We first consider  the case of JWST-ER1,  a strongly lensed galaxy with a
complete Einstein ring and  a
high lensing-to-stellar mass ratio \cite{Dokkum23}. The  lens redshift  is $z_{\mathrm{L}}=1.94$; it is a massive quiescent galaxy with a low star-formation rate and an age of $\tau_\mathrm{L} = 1.9 ^{+0.3}_{-0.6}$ Gyr, implying mainly low-mass stars. Its unit-less age is $t_{\mathrm{L}}=0.62$, according to 
\begin{equation}
t= \left[\Omega_{\mathrm{m}}+(1+z)^{-3/2}(1-\Omega_{\mathrm{m}})\right]^{1/3}.
\label{tz}
\end{equation} 
as derived from (\ref{Jesus}) with   $\Omega_{\mathrm{m}} =0.05$, i.e., assuming  no dark matter. According to the scaling,  $M \sim t$, this implies that   masses in JWST-ER1 are  a fraction 0.62 of the present ones. This shift in mass is  rather  similar to the solution proposed by van Dokkum et al. \cite{Dokkum23}  to solve the discrepancy between the stellar mass ($M_{\star}=1.1 \times 10^{11}$ M$_{\odot}$) and the lensing mass ($M_{\mathrm{Tot}} = 6.5^{+3.7}_{-1.5} \times 10^{11} \; M_{\odot}$).
They  suggested a shift from Chabrier's IMF \cite{Chabrier03}, with few low-mass stars giving   $M_* = 1.1^{+0.2}_{-0.3} \times 10^{11} \; M_{\odot}$, to a very bottom-heavy IMF (``Super-Salpeter IMF''), giving  $M_* = 4.0^{+0.6}_{-0.8} \times 10^{11} \; M_{\odot}$.  The standard Salpeter IMF gives an intermediate value of $M_* = 2.0^{+0.5}_{-0.5} \times 10^{11} \; M_{\odot}$. Figure \ref{imf} illustrates the various  IMF's. 

\hspace{22pt}New results on JWST-ER1 by Mercier et al. \cite{Mercier24}  give a much higher redshift\linebreak 
$z_{\mathrm{S}}= 5.48\pm0.06$ of the source instead of 2.98 by \cite{Dokkum23},  while $z_{\mathrm{L}}=2.02\pm 0.02$ is about the same. The estimated masses are also different; they are, respectively, $M_{\mathrm{Tot}} = (3.66\pm 0.36) \times 10^{11}$ M$_{\odot}$  and a stellar mass of  $ M_*= 1.37^{+0.14}_{-0.11} \times 10^{11}$ M$_{\odot}$, which partially reduces the discrepancy.
Even with the Super-Salpeter assumption, some fraction of dark matter is  necessary, for both studies \cite{Dokkum23,Mercier24}.

\hspace{22pt}Since the deflection angle is free from scale-invariant effects, we  now turn to the mass estimates from photometry. 
The observed luminosity $L_{\mathrm{obs}}$  reflects the  stellar  masses  at the age we see them. For an age  $\tau_\mathrm{L} = 1.9^{+0.3}_{-0.6}$ Gyr assigned by Prospector {(PROSPECTOR is %MDPI: Footnotes should not be used. Information was moved within main text in parenthesis. Please check and confirm.
 a flexible code for inferring stellar population parameters from photometry and spectroscopy spanning UV through IR wavelengths \cite{Johnson21})}, the turnoff mass is around 1.7 $M_{\odot}$ \cite{Maeder09}, in agreement with a remark by \cite{Longair98} that most of the light of  the old  populations in galaxies comes from stellar masses in the range of $[1-2]$ M$_{\odot}$. However, 
a shift in mass by a factor 0.62  for JWST-ER1 implies a big change  in the relative number frequency of stars with respect to the standard case and has great incidence.
\hspace{22pt}\textls[-15]{Let us estimate the effect of such a shift on the average mass--luminosity ratio   $\langle M/L \rangle$ of a galaxy. As an example, we take an initial mass function $\xi(\ln m)$  of the Salpeter form \citep{Salpeter55}:}
\begin{equation}
\xi(\ln m)= \frac{dN}{d\ln \, m}=\frac{dN}{dm} m= A m^{-x}, \quad \mathrm{or} \; \frac{dN}{dm} = A m^{-(1+x)},
\label{dnm}
\end{equation}
with $x=1.35$ (see Figure \ref{imf}). The mass of the galaxy then writes
\begin{equation}
M_{gal}= \int \frac{dN}{dm} m \, dm= A \int m^{-x} dm=\;
\mid \frac{A}{-x+1} ({m}^{-x+1})\mid_{m_{min}}^{m_{max}}\;\simeq A m_{m},
\label{mg}
\end{equation}
where $A=const.$  and  $m_{m}$ the mean mass over the IMF ($m_m= $13.7  M$_{\odot}$ for a mass range from $ 0.01$  to $100$ M$_{\odot}$). If we call $\ell(m)$ the luminosity of a star of mass $m$, the stellar  mass--luminosity relation writes $\ell(m) =B \, m^{\alpha}$, where $B$ is another  constant.  The luminosity of a galaxy is
\begin{equation}
L_{gal}= \int  \ell(m)\frac{dN}{dm} dm=A B \int m^{\alpha}\, m^{-(x+1)}dm = \;
\mid\frac{AB}{\alpha-x} {m}^{\alpha-x}\mid_{m_{min}}^{m_{max}}\;= AB \, m_L ,
\end{equation}
where $m_L$ is the mean mass over the luminosity distribution. Typical values are 1.7--1.8 M$_{\odot}$. Thus, the mean mass--luminosity ratio of a galaxy scales like
\begin{equation}
\left \langle \frac{M}{L} \right \rangle = %\frac{ (\alpha-x)}{B(1-x)} \frac{m_m^{-x+1}}{m_L^{\alpha-x}}= 
\frac{ (\alpha-x)}{B(1-x)} m^{1-x}  \; m^{x-\alpha} . %  \sim \; m_L^{x-\alpha}  ,
%\frac{1}{B} \frac{\alpha-x}{1-x}  m^{1-\alpha}= M$_{\odot}$
\label{LM}
\end{equation}
where $m$ is a representative mass; the exact value is not critical, since  all  masses are reduced by the same factor. The exponents $\alpha$ are  large, hence making $\ell$ a steep function of $m$ (Figure \ref{imf}). We have $\ell  \sim  m^{\alpha}$,
%

%\begin{adjustwidth}{-\extralength}{0cm}
%\centering %% If there is a figure in wide page, please release command \centering
\begin{equation}
 \alpha=3.69,   \hskip 2pt m \in [9 - 2]    \; M_{\odot}  \quad \quad
                                         \alpha=4.56,   \hskip 2pt m \in [2 - 1]    \; M_{\odot} \quad  \quad
                                        \alpha=4.00,   \hskip 2pt m \in [1 - 0.6] \; M_{\odot},
\end{equation}
%\end{adjustwidth}
%
according to \cite{Maeder09}. The  individual  $m/\ell$ ratios are varying  with  power $(1-\alpha)$ of the  mass $m$, while   Equation (\ref{LM}) shows that the factors are different when integrated over the mass spectrum of a galaxy. Now, we can see what is doing a change in the mass by a factor $f=0.62$ (or 0.614 by \cite{Mercier24}). The ratio of the modified overall mass--luminosity ratio  to the standard unmodified case is
\begin{equation}
\frac{\left \langle \frac{M}{L} \right \rangle_f}{\left \langle \frac{M}{L} \right \rangle_{\mathrm{std}} }  \simeq  f^{1-x}  \; f^{x-\alpha} = f^{1-\alpha}.
\label{548}
\end{equation}

\hspace{22pt}For a  change in all masses by a factor $f$,  we obtain the same dependence  as by the individual stellar expression  $(m/\ell) \sim m^{1-\alpha}$. We recall that the  ratio of the total mass from lensing to the stellar mass from photometry  $\frac{M_{\mathrm{Tot}}}{M_*}$ was equal to about 5.9 by van Dokkum \cite{Dokkum23} and about 2.7 by Mercier et al. \cite{Mercier24}. For of factor $f=0.62$  (or 0.614 by \cite{Mercier24})  and $\alpha=4.56$, we obtain a value of  5.48  (or 5.67)  for the   ratio  in Equation (\ref{548}). We obtain, respectively,  4.32  and 2.65 with the redshift by Mercier et al.  for   exponents  $\alpha=4$ and 3 in the M-L relation. Such a value $\alpha=3$ applies if some star formation is still active in JWST-ER1, which is not unlikely since this galaxy is observed close to the peak of star formation in the Universe. Thus,   whatever the exact average $\alpha$-value in JWST-ER1 is,  we may conclude 
 the following:   {\emph{due to the shift in masses, the real scale-invariant M/L ratio  is higher  than the one usually considered, which then leads to too low photometric masses, as currently obtained.}}
In the scale-invariant theory, a standard mass distribution  can  naturally account  for the lensing mass without calling for dark matter. 

In the lens, the stellar masses are smaller, then their luminosity is also smaller. If the M-L relation were linear, there would be no particular effect for the derived masses, see (\ref{548}). But, as $m/\ell$
is much larger for reduced $m$, not accounting that is leading to too low mass estimates.

.

\subsection{The Masses from the Sloan Lens ACS (SLACS) Survey}  \label{SLACS}

\hspace{22pt}To  base our analysis on a larger sample,  we examine the Sloan Lens ACS (SLACS) survey by Bolton \cite{Bolton2008}, which is one of the largest and most homogeneous datasets of galaxy-scale lenses, containing about 100 lenses analyzed by Auger et al. \cite{Auger09}. The distribution of luminous and dark matter is studied to redshift $z \approx 0.5$. It is based on multi-band imaging with ACS, WFC, and NICMOS on the HST, analyzed with a lens model using an isothermal ellipsoid mass distribution, and providing high precision mass measurements.  The mass estimates appear unbiased compared to the estimates from SDSS photometry and are in the range $10^{10.5} < M < 10^{11.8}$ $M_{\odot}$, with a typical statistical error of 0.1 dex. The mean M-L ratio of the sample  is  $\langle M/L \rangle = 8.7$, corresponding to a mean stellar mass of  1.84 $M_{\odot}$, which is not much different  from the case of JWST-ER1. 

\hspace{22pt}The survey provides for each lens the fractions $f^{Chab}_{*,Ein}$ and $f^{Salp}_{*,Ein}$ of the stellar mass within the Einstein radius for the Chabrier and Salpeter IMFs \cite{Chabrier03,Salpeter55}, with respect to the lensing mass. These fractions are independent of any priors from lensing, so that, in some cases, fractions higher than 1 were obtained \cite{Auger09}. The mean stellar mass fraction within the Einstein radius is  $\langle f^{Chab}_{*,Ein}\rangle=0.4$ $\pm 0.1$ for the Chabrier IMF; for the Salpeter's IMF, the mean is $\langle f^{Salp}_{*,Ein}\rangle =0.7 \pm 0.2.$ 
\hspace{22pt}In the sample by \cite{Auger09}, we select  three sub-samples: low redshifts, lenses with $z_{\mathrm{L}} \leq 0.15$; intermediate, with $z$ between  0.15 and 0.30;  higher redshifts, with $z_{\mathrm{L}}  \geq 0.30$. This    is a compromise between the need to have   different  mean $z_{\mathrm{L}}$ and a sufficient number of galaxies in each sub-sample. The results are shown in Table \ref{tab:SLACS}.  We see that the stellar mass fractions are lower  for the higher  $z$-interval: compared to the lowest $z$-interval, the mean differences are  $\Delta f_* = 0.092$ for the Chabrier IMF and  $\Delta f_* = 0.166$ 
for the Salpeter IMF. The values of the stellar mass fractions for the intermediate zone lie in between. These differences are of the same order as the uncertainties. The standard deviations of the different fractions are also given in Table \ref{tab:SLACS}, allowing  tests of significance.

\begin{table}[H]
\caption{Analysis of the SLACS Survey: the mean stellar mass fractions within the Einstein radius, $\langle f^{Chab}_{*,Ein}\rangle$  and  $\langle f^{Salp}_{*,Ein}\rangle$  are given for  Chabrier and Salpeter IMFs respectively as well as the corresponding standard deviations. The column $\langle f^{Salp}_{SIV,Ein} \rangle $  gives the theoretical value  of the stellar mass fraction with Salpeter's IMF obtained after accounting for the scale invariant correction; ideally it should be 1.0.  The last colum gives the number of galaxies in each sub-sample. \label{tab:SLACS}}.
\begin{center} 
%\scriptsize
\begin{tabular}{cccccccc}
\hline
$\mathrm{Redshift \; interval}$ &  $\langle z \rangle $  & $\langle f^{Chab}_{*,Ein}\rangle$  &  $\langle f^{Salp}_{*,Ein}\rangle$ &  
   $\sigma(f^{Chab}_{*,Ein})$   &$\sigma(f^{Salp}_{*,Ein})$ & $\langle f^{Salp}_{SIV,Ein} \rangle $ & $ N $\\
\hline\hline
$0  \leq z  \leq 0.15 $     &  0.114  & 0.449  & 0.792 & 0.110     & 0.197  & 0.95 & 17 \\
$0.15  \leq z  \leq 0.30 $ &  0.218  & 0.419  & 0.745 & 0.053     & 0.130  & 1.04 & 39 \\
$ 0.30 \leq z \leq 0.50   $&  0.383  & 0.357  &  0.626 & 0.079    & 0.136 & 1.07 & 10  \\
\hline
\end{tabular}
\end{center}
\normalsize
\end{table}

\hspace{22pt}Combining  the dispersions for the low and   high $z$-intervals, we obtain $\sigma= 0.135$  for  the difference $ \Delta f_* = 0.449 -0.357= 0.092$  of the  $\langle f^{Chab}_{*,Ein}\rangle$ values. For a normal distribution, this corresponds to a probability of 50.4\% that the difference is significant. For $\langle f^{Salp}_{*,Ein}\rangle$, the difference is $\Delta f_* = 0.792-0.626=0.166$ and  $\sigma=0.239$, giving a probability of 51.2\%. Thus,  the differences are possible, but also  compatible with a random realization. The ages $t$ corresponding to the mean redshifts $z=0.114$, 0.218,  and 0.383 of the samples, respectively, are  $t=0.8591, 0.9113$, and 0.9502  in the dimensionless time scale, again for $\Omega_{\mathrm{m}}=0.05$. According to the modified conservation laws \cite{Maeder17a},  masses are scaling like $t$, which  implies an increase in the mass with decreasing redshifts%Please check that intended meaning has been retained
.  The observations effectively suggest an increase in  the fractions $f^{Chab}_{*,Ein}$ and $f^{Salp}_{*,Ein}$. Let us make a comparison of observations and theory,   even if the statistical uncertainties remain large.

\hspace{22pt}As  for JWST-ER1, the Salpeter IMF gives masses in better agreement with observations than Chabrier's  IMF: the mass estimates from  photometry at $\langle z \rangle =0.114$ give a stellar mass equal to 79.2\% of the  lensing mass. If we account---as  for JWST-ER1---for the shift of the Salpeter IMF resulting from smaller masses, we need to rescale the stellar fraction in the Einstein radius following Equation (\ref{548}). For  the low redshift bin, we obtain today's corresponding value of the stellar mass fraction of $\langle f^{Salp}_{SIV}\rangle = 0.792 \times  0.9502^{-3.56}=  0.95$.  For the intermediate and high reshift bins, we obtain the fractions  $\langle f^{Salp}_{SIV}\rangle$ = 1.04 and 1.07. These  fractions, close to 1.0, mean that, when a shift in mass due to scale invariance is accounted for in absence of dark matter, the stellar masses from photometry  within the Einstein ring for a standard IMF are remarkably close to the total masses from lensing. Thus, it releases or even suppresses the need for dark matter in explaining the discrepancy between stellar and lensing masses. 

% These results are in the line of those for the previous 
%tests.  All the  astronomical observations studied here concern in fact  %the predicted geodesics of General Relativity: the   velocities of %galaxies   in clusters, the rotation curves of  galaxies at low and high %redshifts,
%the wide binaries and finally the lensing which concerns the geodesics %of light rays.
%%%%%%%%%%%%%%%%%%%%%
\section{Conclusions}  
\label{concl}
%%%%%%%%%%%%%%%%%%%%%

\hspace{22pt}The SIV theory rests on a  simple fundamental principle: enlarging the group of invariances that subtend the theory of gravitation, as suggested by Dirac \cite{Dirac73}, by including scale-invariant properties. Effects occur only in low-density ($\varrho < \varrho_{\mathrm{c}}$) Universe models and generally depend  on the inverse of the age of the Universe.
 Even in a low-density Universe, 
 the SIV effects are largely dominated by Newton gravity, while their cumulative contributions may play some role in  the  evolution of locally low-density  structures.  Of note is that the theory contains no ``adjustment parameter''.

\hspace{22pt}There is an impressive convergence of results concerning  several  fundamental problems in astrophysics. They include the following:  (1) %MDPI: List format was revised. Please check and confirm. Same below.
 the observed excesses in velocity for member galaxies in clusters with respect to the visible matter; (2) the relatively flat or shallow slope of the rotation curves in low redshift galaxies; (3) the observed steeper rotation curves of high-redshift galaxies; (4) the omnipresent  relation between dark and baryonic matters  observed in galaxy clusters and  galaxies at different redshifts; (5) the excess of velocity in very wide binary stars with  separations larger than about 3000 kau; (6) the discrepancy between the (larger) masses of galaxies derived from lensing and the (lower) stellar masses obtained from their photometry.

\hspace{22pt}The results (with an appropriate size!) all support the occurrence of  scale-invariant effects  and
throw into  question the need for---or even the very existence of---dark matter%Please check that intended meaning has been retained
. The observations considered here, in addition to those of previous studies  (in particular on the growth of density structures \cite{MaedGueor19}), concern a variety of problems (see also the introduction); they come from a broad range of sources and authors and they  happen in  astronomical objects spanning  a vast range 
of evolutionary stages, with masses from 1 M$_{\odot}$ to about 10$^{14}$  M$_{\odot}$, and spatial scales  from  0.01 pc  to a few  Gpc.
%The dark matter hypothesis appears to lead to some contradictory %results: there should be less dark matter at higher redshifts (steep %rotation curves of high $z$ galaxies), while lensing would suggest the %opposite: with more dark matter needed at higher redshift than at low %redshift according to JWST-ER1 and SLACS. 
In view of the  different  tests performed, it seems unlikely that the scale-invariant theory explains so many observations---and ones which are so diverse---just by chance, especially in the absence of any  fitting  parameter. 

\vspace{6pt}
\authorcontributions{Writing %MDPI: Frédéric Courbin is not mentioned in author contribution section. Please check and revise.
---original draft A.M.; conceptualization---both authors  A.M. and F.C.; formal
analysis---both authors; investigation---both authors; methodology---both authors; validation---both
authors; writing---review and editing---both authors. Both authors have read and agreed to the published version of the manuscript.}

\funding{This research received no external funding.}

\dataavailability{No new data were created or analyzed in this study.}

\acknowledgments{A.M.  expresses his gratitude to James Lequeux and Georges Meynet for their support and encouragement, and  to Vesselin Gueorguiev for constructive collaboration.%Please check that intended meaning has been retained
	.}

\conflictsofinterest{The authors declare no conflict of interest. %Declare conflicts of interest or state ``The authors declare no conflict of interest.'' Authors must identify and declare any personal circumstances or interest that may be perceived as inappropriately influencing the representation or interpretation of reported research results. Any role of the funders in the design of the study; in the collection, analyses or interpretation of data; in the writing of the manuscript; or in the decision to publish the results must be declared in this section. If there is no role, please state ``The funders had no role in the design of the study; in the collection, analyses, or interpretation of data; in the writing of the manuscript; or in the decision to publish the results''.
}

%%%%%%%%%%%%%%%%%%%%%%%%%%%%%%%%%%%%%%%%%%
\begin{adjustwidth}{-\extralength}{0cm}
%\printendnotes[custom] % Un-comment to print a list of endnotes

\reftitle{References}

\PublishersNote{}
\end{adjustwidth}
\end{document}